\pdfoutput=1
\documentclass[%
reprint,
superscriptaddress,
showkeys,
twoside,
amsmath,
amssymb,
aps,
prd,
]{revtex4-2}
\usepackage[english]{babel}
\usepackage[utf8]{inputenc}
\usepackage[T1]{fontenc}

\usepackage{afterpage}

\usepackage{graphicx}
\graphicspath{{figs/}}

\usepackage{booktabs}

\usepackage[pdftex, pdftitle={Article}, pdfauthor={Author}, hidelinks]{hyperref}

\usepackage{journals}

\usepackage{dcolumn}

\usepackage[capitalize]{cleveref}

\usepackage{bm}

\usepackage{xcolor}

\usepackage[]{todonotes}

\usepackage{siunitx}

\usepackage{subfig}

\usepackage{ctable}
\usepackage{multirow}
\usepackage{array}
\newcolumntype{?}{!{\vrule width 2pt}}

\usepackage{soul}
\usepackage{cleveref}

\usepackage{placeins}

\newcolumntype{d}[1]{D{.}{.}{#1}}

\begin{document}

\begin{abstract}
We present a novel universal relation for binary neutron star mergers with long-lived neutron star remnants: inspired by recent work based on numerical relativity simulations, we propose a novel approach using perturbative calculations that allow us to relate the pre-merger neutron star binary tidal deformability to the effective compactness of the post-merger remnant. 
Our results allow for the prediction of the stellar parameters of a long-lived remnant neutron star from the study of gravitational waves emitted during the pre-merger phase.
\end{abstract}

\title{Universal Relations for Binary Neutron Star Mergers with Long-lived Remnants}

\author{Praveen Manoharan}
	\email{praveen.manoharan@uni-tuebingen.de}
	\affiliation{Theoretical Astrophysics, IAAT, University of T\"ubingen, 72076 T\"ubingen, Germany}
\author{Christian J. Kr\"uger}
    \email{christian.krueger@tat.uni-tuebingen.de}
    \affiliation{Theoretical Astrophysics, IAAT, University of T\"ubingen, 72076 T\"ubingen, Germany}
\author{Kostas D. Kokkotas}
    \email{kostas.kokkotas@uni-tuebingen.de}
    \affiliation{Theoretical Astrophysics, IAAT, University of T\"ubingen, 72076 T\"ubingen, Germany}
    
\date{\today}

\newcommand{\blue}[1]{{\color{blue}#1}}


\maketitle

\section{Introduction}\label{sec:introduction}
The successful detection of gravitational waves from binary neutron star (BNS) mergers through the LIGO-VIRGO detectors~\cite{2017PhRvL.119p1101A,2020ApJ...892L...3A}
has opened a new avenue into probing and understanding the structure of neutron stars: through the observation of these mergers, constraints can be put on the radius, maximum
TOV mass, and tidal deformability of neutron stars~\cite{1996PhRvL..77.4134A,2001MNRAS.320..307K,2018ApJ...852L..25R,2018PhRvL.120z1103M,2020PhRvL.125n1103B}, which will allow us to uncover their true equation of state (EoS).

An important tool for this task are EoS independent -- or \emph{(approximately) universal} -- relations that allow for the inference of neutron star bulk parameters
through information extracted from gravitational waves (and/or electromagnetic observations). Inspired by earlier work on such universal relations for single neutron stars~\cite{1998MNRAS.299.1059A,2004PhRvD..70l4015B,2017PhR...681....1Y},
the last five years have given rise to universal relations for BNS: they relate the pre-merger neutron stars to the early post-merger remnant, and have been developed using numerical
relativity simulations~\cite{2015PhRvL.115i1101B,2016PhRvD..93l4051R,2020PhRvD.101h4006K}.

These works have primarily focused on relating the tidal deformability of the pre-merger stars, which impact the dynamics of the pre-merger gravitational waves at leading order through the 
the \emph{binary tidal deformability} $\tilde\Lambda$~\cite{2008PhRvD..77b1502F,2014PhRvL.112j1101F}, to various stellar parameters of the early remnant. 
More recently, Kiuchi et al.~\cite{2020PhRvD.101h4006K} have shown that existing universal relations for BNS mergers suffer from systematic 
errors caused by, e.g., only considering pre-merger stars with close to equal masses (a common assumption that had been called into question after the observation of \texttt{GW170817}~\cite{2017PhRvL.119p1101A}). 
Allowing for a wider range of mass ratios than in previous works, they 
propose alternative relations between $\tilde\Lambda$ and post-merger remnant parameters with high accuracy (with maximum relative error at the order of $10^{-2}$).

Recently, Vretinaris et al.~\cite{2020PhRvD.101h4039V} also investigated empirical relations for BNS mergers based on the extensive coRE data set~\cite{dietrich2018core}
of numerical relativity gravitational wave simulations. Covering a wide range of mass ratios, they find an extensive set of universal relations involving the 
various peak frequencies of the post-merger gravitational wave signal, involving, e.g., the chirp mass and characteristic radius of a 1.6 $M_\odot$ neutron star. 
In particular, they also find universal relations between the binary tidal deformability and the primary $f$-mode frequency of the post-merger signal
(as in~\cite{2020PhRvD.101h4006K}), however, this time involving the chirp mass of the BNS.

The investigation of a wider range of the BNS parameter space through numerical relativity simulations, however, remains limited due to their high computational cost. In this paper, we thus
propose a novel approach to developing universal relations for BNS mergers using perturbative calculations, assuming that the merger results in a long-lived remnant neutron star: 
guided by general constraints on, e.g., the total mass and angular momentum of the pre- and post-merger phases of a BNS merger obtained through numerical relativity simulations, 
we can individually treat the neutron stars in these phases and compare their properties to obtain new relations. This approach is primarily enabled by recent work, from two of us, on
computing the $f$-mode frequency for fast rotating neutron stars without approximation~\cite{2020PhRvL.125k1106K,2020PhRvD.102f4026K}.

Inspired by the universal relation between the binary tidal deformability of the BNS and the stable, co-rotating $f$-mode frequency $\sigma^s$ of the early, differentially rotating remnant proposed in~\cite{2020PhRvD.101h4006K}, we, in a first step, 
derive a similar relation for a potentially long-lived, uniformly rotating remnant: the relation takes the form
\begin{equation}
\log_{10} \hat\sigma^s = a(q) \cdot \tilde \Lambda^{\frac{1}{5}} + b(q)
\end{equation}
where $\hat \sigma^s = \frac{M}{M_\odot}\frac{\sigma^s}{\si{k Hz}}$ is the normalized co-rotating $f$-mode frequency, and $q = \frac{M_1}{M_2} \leq 1$ the gravitational mass ratio of the pre-merger stars. For rapidly rotating, long-lived remnants (with rotation frequency $\bar\Omega \geq 800 \si{Hz}$), this relation achieves an average relative error of $1.3\%$.

We also derive a relation for the potentially unstable, counter-rotating $f$-mode frequency of the long-lived remnant, presenting the possibility of predicting the onset of the Chandrasekhar-Friedman-Schutz (CFS) instability~\cite{1970PhRvL..24..611C,1978ApJ...221..937F}.

Combining these results with a universal relation for fast rotating neutron stars we put forward in~\cite{2020PhRvL.125k1106K} between the stable, co-rotating $f$-mode frequency and the \emph{effective compactness} $\eta = \sqrt{\bar M^3/I_{45}}$, where $\bar M = M/M_\odot$ is the normalized, gravitational mass of the neutron star and $I_{45} = I/10^{45} \si{g cm^2}$ its normalized quadrupole moment, 
we also derive a combined relation of the form
\begin{equation}
\eta = \frac{10^{a(q) \cdot \tilde \Lambda^{\frac{1}{5}} + b(q)} - \left(c_1 + c_2 \hat \Omega + c_3 \hat \Omega^2\right)}{d_1 + d_3 \hat \Omega}
\end{equation}
that relates the pre-merger binary tidal deformability of the BNS with the effective compactness of the long-lived remnant. For rapidly rotating remnants, this relation achieves an average relative error of $2.4\%$.

Finally, by directly relating these quantities without going via the $f$-mode, we obtain a universal relation of the form
\begin{equation}
\log\left[\bar M^5 \eta\right] = a(q) \left(\bar M^{5}\tilde\Lambda^{-\frac{1}{5}}\right)^2 + b(q) \bar M^{5}\tilde\Lambda^{-\frac{1}{5}}+ c(q),
\label{eq:direct}
\end{equation}
This relation achieves improved accuracy, reaching an average relative error of $\sim 1.5\%$ for remnants with any rotation frequency.

We also consider a direct relation between the binary tidal deformability and the compactness $C = \frac{M}{R}$ of the long-lived remnant. 
Such a relation would allow the direct estimation of the remnant's radius $R$ using independent estimates of its gravitational mass. We propose a relation of the form
\begin{equation}
\bar M^5 C = a(q) \bar M^{5} \tilde \Lambda^{-\frac{1}{5}} + b(q)
\end{equation}
which, however, only achieves an accuracy an order of magnitude worse than for the effective compactness relation, reaching an average relative error of $\sim 8.8\%$.

While our approach allows for a wide range of parameters for the long-lived remnant, a number of works exist that put constraints on them: Radice et al.~\cite{2018MNRAS.481.3670R}, for instance, 
constrain the remnant's rotation period based on its baryon mass. We adopt this constraint and investigate potential improvements to the above relations. Our findings show that this constraint 
generally improves the accuracy of our relations, but only slightly. 

Finally, we also investigate the impact of a non-negligible baryon mass loss going from the pre-merger stars to the long-lived remnants, as e.g. suggested in~\cite{2018MNRAS.481.3670R}. For bounded mass losses of up to 
$0.2 M_\odot$, we find that a) the mass loss does not preclude the construction of new best fits of our relations with essentially the same accuracy as before, and b) even our original relations (without mass loss) can be
applied to BNS mergers with mass lass with only a small increase in error.

The results presented in this paper represent a first step towards finding universal relations between the pre-merger neutron stars and the potential long-lived remnant 
of a BNS merger using perturbative calculations. Our approach can be freely extended to e.g. hot EoSs, phase transitions, as well as differential rotation for the remnant to, e.g., cover earlier parts of the post-merger phase. 

\medskip
\noindent
\textbf{Outline.}
We begin by outlining our approach and the methods used for our computations in Section~\ref{sec:methods}. We then investigate the relation between the pre-merger tidal deformability and the $f$-mode frequency of the remnant using perturbative calculations in Section~\ref{sec:f-mode}. Combining these relations with the results presented in~\cite{2020PhRvL.125k1106K}, we obtain the combined relation between the pre-merger tidal deformability and remnant effective compactness. In Section~\ref{sec:direct} we then present our novel, direct relation between these quantities that achieves improved accuracy, and also introduce the compactness relation. We then discuss the impact of introducing constraints on the remnant's rotation rate in Section~\ref{sec:constrained}, and the impact of a non-negligible mass loss on our relations in Section~\ref{app:mass_loss}. Finally, we conclude our work and discuss potential directions for future work in Section~\ref{sec:conclusion}. 

Note that, throughout the paper, we will assume geometrized units in which $G = c = 1$.
\section{Methodology}\label{sec:methods}
In this section, we briefly describe our model for the BNS merger and how we determine the desired properties of the pre-merger
and post-merger neutron stars.

\subsection{Binary Neutron Star Systems}\label{sec:binary}
For our analysis, we propose a simplified model of a BNS merger: we consider a binary of irrotational neutron stars with gravitational masses $M_1$ and $M_2$
and mass ratio $q = \frac{M_1}{M_2} \leq 1$. The merger of these two stars results in a long-lived remnant of \emph{baryon mass} $M_b = M_{b,1} + M_{b,2}$, where $M_{b,1}$ and $M_{b,2}$ are the baryon masses of the
respective pre-merger stars. We denote the remnant's gravitational mass with $M$.

For the sake of simplicity, we assume the baryon mass (or rest mass) loss during the merger and spin-down, i.e., the transition from the early, differentially rotating remnant to the long-lived, uniformly rotating remnant, to be negligible: 
numerical relativity simulations have shown that the baryon mass loss 
during the merger is typically small, at the order of $10^{-4} - 10^{-2} 
M_\odot$~\cite{2013ApJ...773...78B,2013PhRvD..87b4001H,2016PhRvD..93l4046S,2018ApJ...869..130R}. 
While the baryon mass loss during the spin-down of the remnant is potentially larger, it is also not yet fully understood: 
Radice et al.~\cite{2018MNRAS.481.3670R} give an estimate of up to $0.2 M_\odot$ baryon mass that is ejected through a combination of viscous processes and neutrino emission during this period, however also state
that the details of these processes need to be further investigated. As such, we first remain with the assumption of a negligible baryon mass loss while we develop our relations. 
We then quantify the the impact of a non-negligible baryon mass loss in Section~\ref{app:mass_loss}.

We generally assume the long-lived remnant to be a massive neutron star with gravitational mass $M \geq 2 M_\odot$, up to the
maximum mass supported under uniform rotation, 
which, depending on the EoS, reaches values around $2.3-3.3 M_\odot$. 

To study the dependence of our relations on the rotation rate of the remnant, we allow angular rotation rates $\Omega$ between $0$ (non-rotating) 
and the Kepler mass-shedding limit $\Omega_K$ (maximally rotating). However, we will later introduce constraints on $\Omega$ in
an attempt to improve the accuracy of our relations under physical conditions. These constraints will be based on estimates of the expected spin of 
the post-merger remnant put forward in.~\cite{2018MNRAS.481.3670R}. 

For a wide range of the three primary parameters $M, q$ and $\Omega$, and a selection of phenomenological EoSs, we then compute the binary tidal deformability $\tilde\Lambda$ 
of the pre-merger objects, and the $f$-mode frequency and effective compactness of the long-lived remnant, as described below.

Implicitly, we assume that the long-lived remnant shares the same \emph{cold} EoS as the pre-merger neutron stars and 
that the initially differential rotation of the remnant has already been driven to uniform rotation by viscous processes.
Numerical relativity simulations show that these conditions are usually achieved at a cooling timescale of $\sim 2-3 \si{s}$~\cite{2018MNRAS.481.3670R}.

We also do not consider the possibility of a phase transition between the pre-merger EoS and remnant EoS~\cite{2019PhRvL.122f1102B}, i.e. 
the EoS remains the same throughout the merger. As discussed by Nandi and Pal~\cite{2021EPJST.tmp....5N}, phase transitions from purely hadronic matter to quark matter 
can be useful to address issues with reconciling tidal deformability upper bounds derived from the observation of \texttt{GW170817}~\cite{2017PhRvL.119p1101A} (implying a softer EoS), 
with the maximum mass lower bound derived from the observation of the millisecond pulsar \texttt{J0740+6620}~\cite{2020NatAs...4...72C} (implying a stiffer EoS).

Such phase transitions, however, change the properties of neutron stars. In particular, they lead to, generally, smaller tidal deformabilities~\cite{2019PhRvD..99h3014H} compared to the purely
hadronic case. As shown in~\cite{2021EPJST.tmp....5N} using the well-known I-Love-Q relation~\cite{2017PhR...681....1Y}, 
this also affects the accuracy of universal relations that rely on the tidal deformability. 
As such, extending our work to include EoSs with phase transitions will be an important direction for further work.
\subsection{Equilibrium Models}\label{sec:equi}
For the equilibrium model of the non-rotating neutron stars we consider a metric of the form
\begin{equation}
ds^2 = -e^{\nu(r)} dt^2 + e^{\lambda(r)} dr^2 + r^2 d\Omega^2
\end{equation}
and solve the TOV equations assuming a perfect fluid (for a more detailed description, we refer to Appendix~\ref{app:equi}).

For the EoS, we utilize piecewise-polytropic 
approximations, proposed by Read et al.~\cite{2009PhRvD..79l4032R}, of the five realistic EoSs \texttt{SLy}~\cite{2001A&A...380..151D}, \texttt{WFF1}~\cite{1988PhRvC..38.1010W}, \texttt{APR4}~\cite{1998PhRvC..58.1804A}, \texttt{H4}~\cite{2006PhRvD..73b4021L} and \texttt{MS1}~\cite{1996APS..MAY..E704M}.

These piecewise-polytropic approximations are defined as follows: for a given set of rest-mass density thresholds $\rho_i$, the
pressure-density relations of these EoSs are given by
\begin{equation}
p(\rho) = K_i \rho^{\Gamma_i}\quad \rho_{i-1} \leq \rho < \rho_i
\label{eq:eos}
\end{equation}
where $\Gamma_i$ is the adiabatic index, and $K_i$ the proportionality constant specific to the given EoS. All of these constants are, in particular, chosen in such a way that continuity of the pressure
is ensured at each rest-mass density threshold $\rho_i$.

For these piecewise-polytropic EoSs, the energy density $\epsilon$ and rest mass density $\rho$ are related by
\begin{equation}
\epsilon(\rho) = (1+a_i)\rho + \frac{K_i}{\Gamma_i -1}\rho^{\Gamma_i}
\label{eq:eden_rhoc}
\end{equation}
where the $a_i$ are EoS specific constants. The values for the constants $K_i$, $\Gamma_i$ and $a_i$ for the five EoS that we use in 
this paper are given in tabulated form in~\cite[Table III]{2009PhRvD..79l4032R}.

\begin{figure}
\centering
\includegraphics[width=0.5\textwidth]{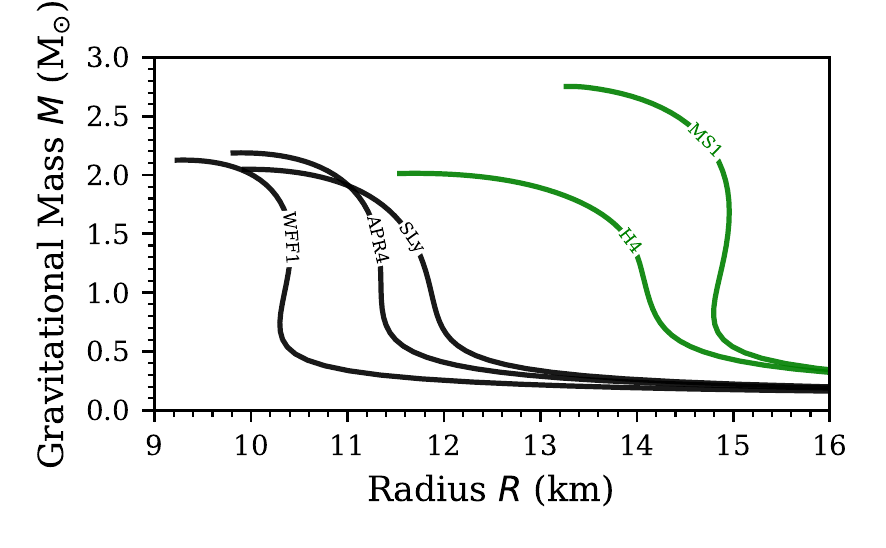}
\caption{The mass-radius relation for non-rotating stars of the five phenomenological EoSs considered in this paper. EoSs that we consider \emph{soft} are indicated in black, and those we consider \emph{stiff} in green.}
\label{fig:mass-radius}
\end{figure}

We integrate the TOV equations using the ODE solver \texttt{solve\_ivp} implemented in the \texttt{Python} package \texttt{Scipy}, which internally implements the explicit 5th-order Runge-Kutta method.
The resulting mass--radius relations for the five EoSs considered in this paper are given in Figure~\ref{fig:mass-radius}. To help discussions in the following sections, we distinguish EoSs that we consider \emph{soft} (WFF1, APR4 and SLy) and \emph{stiff} (H4 and MS1) by color.

\subsection{Tidal Deformability}\label{sec:Love}
To compute the tidal deformability of the irrotational pre-merger neutron stars, we follow the formalism presented by Hinderer~\cite{2009ApJ...697..964H}. 
Alternative, but more general formulations can be found in Damour and Nagar~\cite{2009PhRvD..80h4035D}, and Binnington and Poisson~\cite{2009PhRvD..80h4018B}.

The tidal Love number $k_2$ of an irrotational neutron star with gravitational mass $m$ and radius $R$ is given by
\begin{equation}
\begin{split}
k_2 =& \frac{8 C^5}{5}\left(1-2C^2\right)\left[2 + 2C(y -1) - y\right]\\
& \times \left\{2 C \left[6 - 3y + 3C\left(5y-8\right)\right] \right.\\
&\qquad \left. + 4C^3\left[13-11y+C(3y-2)+2C^2(1+y)\right] \right.\\
&\qquad \left. + 3(1-2C^2)\left[2-y+2C(y-1)\right]\ln(1-2C)\right\}^{-1}
\end{split}
\end{equation}
where $C = \frac{m}{R}$ is the compactness of the neutron star, and $y = R \frac{H'(R)}{H(R)}$. Here, the function $H$ is derived from the $rr$ and $tt$ components of the static, polar perturbations of the neutron star~\cite{1967ApJ...149..591T}, and is determined by (Hinderer~\cite[Eqn. 15]{2008ApJ...677.1216H})
\begin{equation}
\begin{split}
& H'' + H'\left[\frac{2}{r} + e^{\lambda(r)} \left(\frac{2 m(r)}{r^2} + 4 \pi r(p-\epsilon)\right)\right] \\
&+ H\left[-\frac{6e^{\lambda(r)}}{r^2} + 4 \pi e^{\lambda(r)}\left(5 \epsilon + 9p + \frac{\epsilon + p}{d p/ d\epsilon}\right) - \nu'(r)^2\right] \\
&= 0
\end{split}
\label{eq:main}
\end{equation}	
Our task will thus be to integrate Equation~\eqref{eq:main} from $r = 0$ to $R$ to obtain the values for $H(R)$ and $H'(R)$. As initial values for $H$ and $H'$ we follow
the suggestion in~\cite{2009PhRvD..80h4035D} of $H(r_0) = r_0^2$ and $H'(r_0) = 2 r_0$ for some small initial radius $r_0 \sim 10^{-6}$.

\begin{figure}[t]
\includegraphics[width=0.5\textwidth]{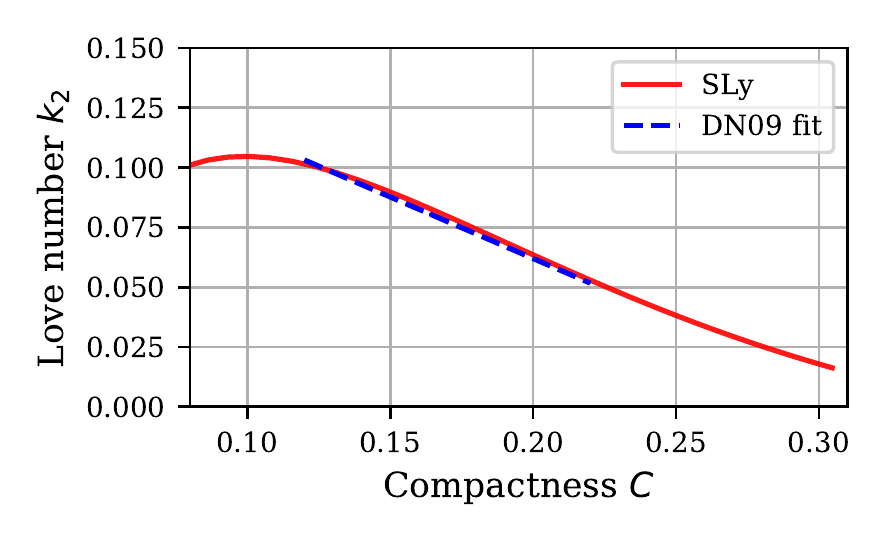}
\caption{The relation between the compactness $C$ and Love number $k_2$ for the EoS SLy, together with the linear fit proposed in~\cite{2009PhRvD..80h4035D}.}
\label{fig:love}
\end{figure}

Exemplarily, we illustrate the resulting Love numbers for the EoS SLy depending on the compactness $C$, 
together with the linear fit proposed by Damour and Nagar~\cite[Equation (116)]{2009PhRvD..80h4035D} (for $0.12 \leq C \leq 0.22$) in Figure~\ref{fig:love}.

In the following analyses, we are mainly interested in the \emph{binary tidal deformability} $\tilde\Lambda$~\cite{2014PhRvL.112j1101F} of a BNS before the merger: for a BNS of
mass ratio $q = \frac{M_1}{M_2} < 1$, $\tilde\Lambda$ is given by~\cite{2014PhRvL.112j1101F}
\begin{equation}
\begin{split}
\tilde \Lambda =& \frac{8}{13}\left[\left(1 + 7 \zeta - 31 \zeta^2\right)\left(\hat\lambda_{1} + \hat\lambda_{2}\right)\right.\\
& - \left.\sqrt{1 - 4 \zeta}\left(1 + 9 \zeta - 11 \zeta^2 \right)\left(\hat\lambda_{1} - \hat\lambda_{2}\right)\right]
\end{split}
\end{equation}
where 
\begin{equation}
\zeta = \frac{M_1 M_2}{(M_1 + M_2)^2 } = \frac{q}{(1+q)^2}
\end{equation}
is the \emph{symmetric mass ratio} of the BNS, and
\begin{equation}
\hat \lambda_i = \frac{2}{3} \left(\frac{R_i}{M_i}\right)^5 k_{2,i}
\end{equation}
the \emph{dimensionless tidal deformability} of the $i$-th pre-merger star, with $k_{2,i}$ being its Love number. Our primary goal will be to relate $\tilde\Lambda$ to properties of the post-merger remnant. 

\subsection{Post-merger Remnant}\label{sec:remnant}
As described above, we assume the long-lived remnant of the merger to be a uniformly rotating, massive neutron star with gravitational mass $M \geq 2 M_\odot$ and angular rotation rate $\Omega$, or rotation frequency $\bar \Omega = \frac{\Omega}{2 \pi}$. 
For the stellar parameters of these remnants, we rely on the same data set of rotating neutron stars on which the results in~\cite{2020PhRvL.125k1106K,2020PhRvD.102f4026K}
are also based: there, we used the \texttt{rns}~\cite{1995ApJ...444..306S,1998A&AS..132..431N,RNS} code to create the rotating equilibrium models for the same EoSs that we also consider in this paper, and computed the $f$-mode frequencies of fast rotating neutron stars
without approximation.

Now, for rotating neutron stars, a splitting of the $f$-mode frequency occurs, caused by the $f$-mode oscillations traveling pro- or retrograde to the rotation of the neutron star~\cite{2000CAS....36.....T,2008PhRvD..78f4063G,2017LRR....20....7P}. The counter-rotating mode can pass the zero frequency threshold, reaching negative frequencies and leading to the well-known Chandrasekhar-Friedman-Schutz (CFS) instability~\cite{1970PhRvL..24..611C,1978ApJ...221..937F}. As such, the counter-rotating mode is also referred to as the \emph{potentially unstable} mode, with frequency $\sigma^u$, whereas the co-rotating mode is referred to as the \emph{stable} mode, with frequency $\sigma^s$. From the data set described above, we are able to take the frequency of both the stable, co-rotating $f$-mode 
as well as the potentially unstable, counter-rotating $f$-mode. 

Inspired by universal relations between the pre-merger binary tidal deformability $\tilde\Lambda$ of the BNS and the peak-frequency of the early post-merger signal (which corresponds to the co-rotating $f$-mode of the early remnant) presented in~\cite{2020PhRvD.101h4006K}, we consider the relation between $\tilde\Lambda$ and, both, the co- and counter-rotating $f$-modes of the long-lived remnant.
Combining these results with the universal relation between the $f$-mode frequency and the effective compactness for rotating neutron stars put forward  in~\cite{2020PhRvL.125k1106K}, we are finally able to perform a comprehensive analysis of the relation between pre-merger tidal deformability and remnant properties.

While our initial results are general, and depend on the mass ratio of the pre-merger stars and rotation rate of the remnant, we also consider constraints on these parameters to further improve the accuracy
of our results. To this end, we consider the universal relation for the rotation period of the long-lived remnant put forward by Radice et al.~\cite{2018MNRAS.481.3670R}
\begin{equation}
P = \left[a\left(\frac{M_b}{M_\odot} - 2.5\right) +b\right] \si{ms}
\label{eq:rot_constraint}
\end{equation}
where $a$ and $b$ are EoS specific coefficients, and $M_b$ is the baryon mass of the long-lived remnant. This ansatz is presented as a good approximation for all remnants with a baryon mass of at least $M_b \geq 2 M_\odot$, which also fits our data as we exclusively consider long-lived remnants with a gravitational mass of at least $M \geq 2 M_\odot$.

\section{\texorpdfstring{$f$}{f}-mode Relations for Binary Neutron Stars}\label{sec:f-mode}
Here, we compute the binary tidal deformability $\tilde\Lambda$ of a BNS characterized by its total gravitational mass $M$ and mass ratio $q$,
and compare it to the $f$-mode frequency of the long-lived remnant, as computed in~\cite{2020PhRvL.125k1106K}. 

\subsection{Co-rotating \texorpdfstring{$f$}{f}-mode}
Inspired by the universal relation put forward by Kiuchi et al.~\cite{2020PhRvD.101h4006K} for the early post-merger remnant, we begin by considering the \emph{normalized and dimensionless co-rotating $f$-mode frequency}
\begin{equation}
\hat \sigma^s = \frac{M}{M_\odot}\frac{\sigma^s}{\si{k Hz}}
\end{equation}
of the long-lived remnant at different angular rotation rates $\Omega$. As we mentioned earlier, the superscript $s$ is used to indicate the \emph{stable} (i.e. co-rotating) branch of the $f$-mode for rotating stars.

Without further constraints, this does not immediately lead to a universal relation. In fact, we observe a clear distinction between the soft and stiff EoSs (cf. Section~\ref{sec:equi}), and stars with low angular rotation rates also do not seem to fit very well into a single relation. 

To remedy these issues, we add two constraints: first, since the EoSs used in~\cite{2020PhRvD.101h4006K} correspond mostly to the soft EoSs considered here, we, for now, ignore the stiff EoSs. 
Second, as we expect the remnant to be fast rotating, we require it to have a rotation frequency $\bar \Omega = \Omega / 2\pi$ above a minimum rotation frequency threshold $\bar \Omega_{\mathrm{thr}}$. While
we considered a range of values for $\bar \Omega_{\mathrm{thr}}$ in our analyses, we here exemplarily discuss the result for $\bar \Omega_{\mathrm{thr}} = 800 \si{Hz}$. The corresponding figures for $\bar \Omega_{\mathrm{thr}} \in \{400, 1200\} \si{Hz}$ can be found in Appendix~\ref{app:combined}.

\begin{figure}[t]
\includegraphics[width=0.5\textwidth]{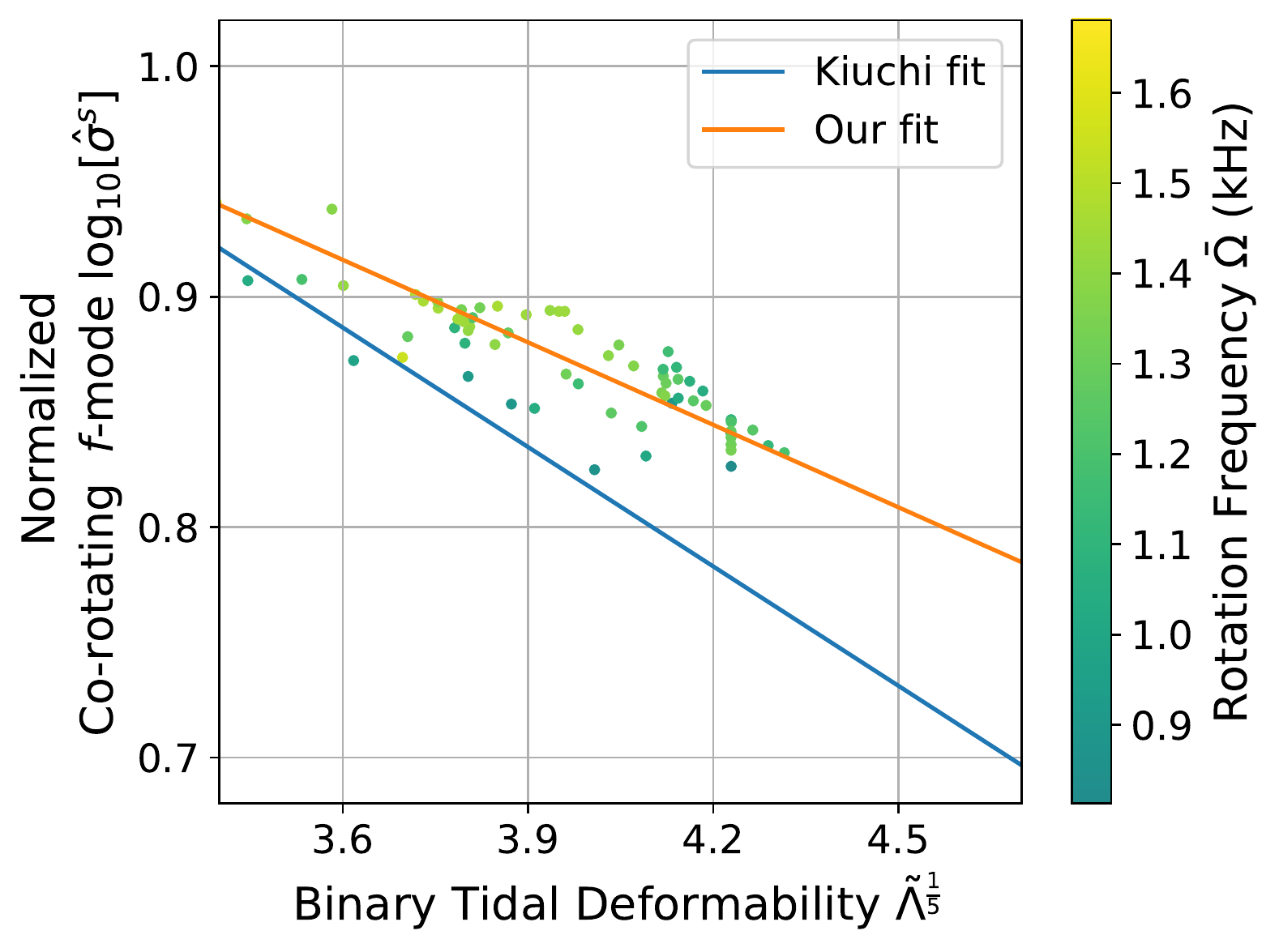}
\caption{The relation between the binary tidal deformability $\tilde\Lambda$ and the (co-rotating) $f$-mode frequency~$\sigma$ for neutron star binaries with mass ratio $q=1$, considering rotation frequencies $\bar \Omega \geq 800 \si{Hz}$. The relation by Kiuchi et al.~\cite{2020PhRvD.101h4006K} was developed for the young post-merger neutron star with hot EoS, causing the observed differences.}
\label{fig:lambda-f}
\end{figure}

The relation that includes the linear fit for stars with soft EoSs (i.e. APR4, SLy and WFF1) and rotation frequencies $\bar \Omega \geq 800 \si{Hz}$ is shown in Figure~\ref{fig:lambda-f}. 
We can clearly observe a deviation from the relation put forward in~\cite{2020PhRvD.101h4006K} that is caused by us considering uniformly rotating neutron stars with cold EoSs (for a long-lived remnant), while the numerical relativity simulations performed in~\cite{2020PhRvD.101h4006K} include differential rotation and thermal corrections for the early post-merger remnants with hot EoSs. However, we are still able to reproduce the same functional form of the relation between $\tilde\Lambda$ and $\hat \sigma$.

Generalizing this linear fit for arbitrary pre-merger mass ratios $q$, we obtain a relation of the form
\begin{equation}
\log_{10} \hat\sigma^s = a(q) \cdot \tilde \Lambda^{\frac{1}{5}} + b(q)
\label{eq:lin_fit}
\end{equation}
where 
\begin{equation}
a(q) = a_2 q^2 + a_1 q + a_0
\end{equation} 
and 
\begin{equation}
b(q) = b_2 q^2 + b_1 q + b_0
\end{equation}

\begin{table*}
\sisetup{round-mode=places,round-precision=3,scientific-notation=true}
\centering
\setlength\tabcolsep{1.5ex}
\renewcommand{\arraystretch}{1.4}
\begin{tabular}{c ? c | c | c | c | c | c ? c | c}
 $\bar \Omega_{\mathrm{thr}} [\si{Hz}]$ & $a_2$ & $a_1$ & $a_0$ & $b_2$ & $b_1$ & $b_0$ & $\overline{E}$ & $\bar e$\\
\specialrule{.2em}{.1em}{.1em} 
400 & \num{0.032687010581885494} & \num{-0.06416109678446326} & \num{-0.08036292462856116} & \num{-0.07673039868157629} & \num{0.15071220409350936} & \num{1.2341755684363536} & \num[scientific-notation=false]{0.027125729335891996} & \num[scientific-notation=false]{0.022907970070006306}  \\
800 & \num{0.03220150630784713} & \num{-0.06317779822336232} & \num{-0.08832940983585191} & \num{-0.07107281083410977} & \num{0.13951282042913238} & \num{1.2770912973520199} & \num[scientific-notation=false]{0.015588717672063497}& \num[scientific-notation=false]{0.013059787036421931}  \\
1200 & \num{0.03347917217676552} & \num{-0.06566164152815367} & \num{-0.09556950169164023} & \num{-0.0710427529003803} & \num{0.13937625091579767} & \num{1.3131969557691394} & \num[scientific-notation=false]{0.012176185263345289}& \num[scientific-notation=false]{0.010911838589497852}  \\
\end{tabular}
\caption[Coefficients for Relation between Binary Tidal Deformability and Co-rotating $f$-mode, with Errors]{Coefficients of the relation for the co-rotating $f$-mode in Equation~\eqref{eq:lin_fit} for different rotation frequency thresholds $\bar\Omega_{\mathrm{thr}}$, together with the RMSE and average relative error. The plot for $\bar \Omega_{\mathrm{thr}} = 800 \si{Hz}$ is shown in Figure~\ref{fig:lambda-f}.}
\label{tab:q-coeff}
\end{table*}
The exact values of these coefficients for different rotation frequency thresholds are given in Table~\ref{tab:q-coeff}. We also give the RMSE $\bar E$ and the average relative error $\bar e$ of the estimated values for $\hat \sigma^s$ through Equation~\eqref{eq:lin_fit} compared to the original value (only considering stars with rotation frequency $\bar \Omega \geq \bar\Omega_{\mathrm{thr}}$). As one would expect, the higher the rotation frequency threshold is chosen, the more accurate the linear fit becomes. For instance, with $\bar\Omega_{\mathrm{thr}} = 800 \si{Hz}$, our linear fit achieves an RMSE of $0.016$, and an average relative error of $\sim 1.3\%$.

\subsection{Counter-rotating \texorpdfstring{$f$}{f}-mode}
Since we have access to the counter-rotating $f$-mode of the rotating remnant neutron star, we also investigate the relation between the binary tidal deformability and the \emph{normalized and dimensionless counter-rotating $f$-mode frequency}
\begin{equation}
\hat \sigma^u = \frac{M}{M_\odot} \frac{\sigma^u}{\si{k Hz}}.
\end{equation}
The superscript $u$ is used to indicate the \emph{potentially (CFS) unstable} (i.e. counter-rotating) branch of the $f$-mode for rotating stars.
We again perform our analyses for different rotation frequencies $\bar \Omega$ and for different mass ratios $q$. We also continue to only consider the soft EoSs (APR4, WFF1 and SLy), as before.

Since the counter-rotating $f$-mode frequency can reach negative values, we cannot straightforwardly adapt
the same form for our relation as above due to the logarithm used there. While there are several approaches one could take to alleviate this issue (e.g. ignore negative frequencies, or only consider the absolute value of the frequency),
our evaluations have shown that the best results are achieved when one attempts a simple linear relation.

\begin{figure}[t]
\centering
\includegraphics[width=0.45\textwidth]{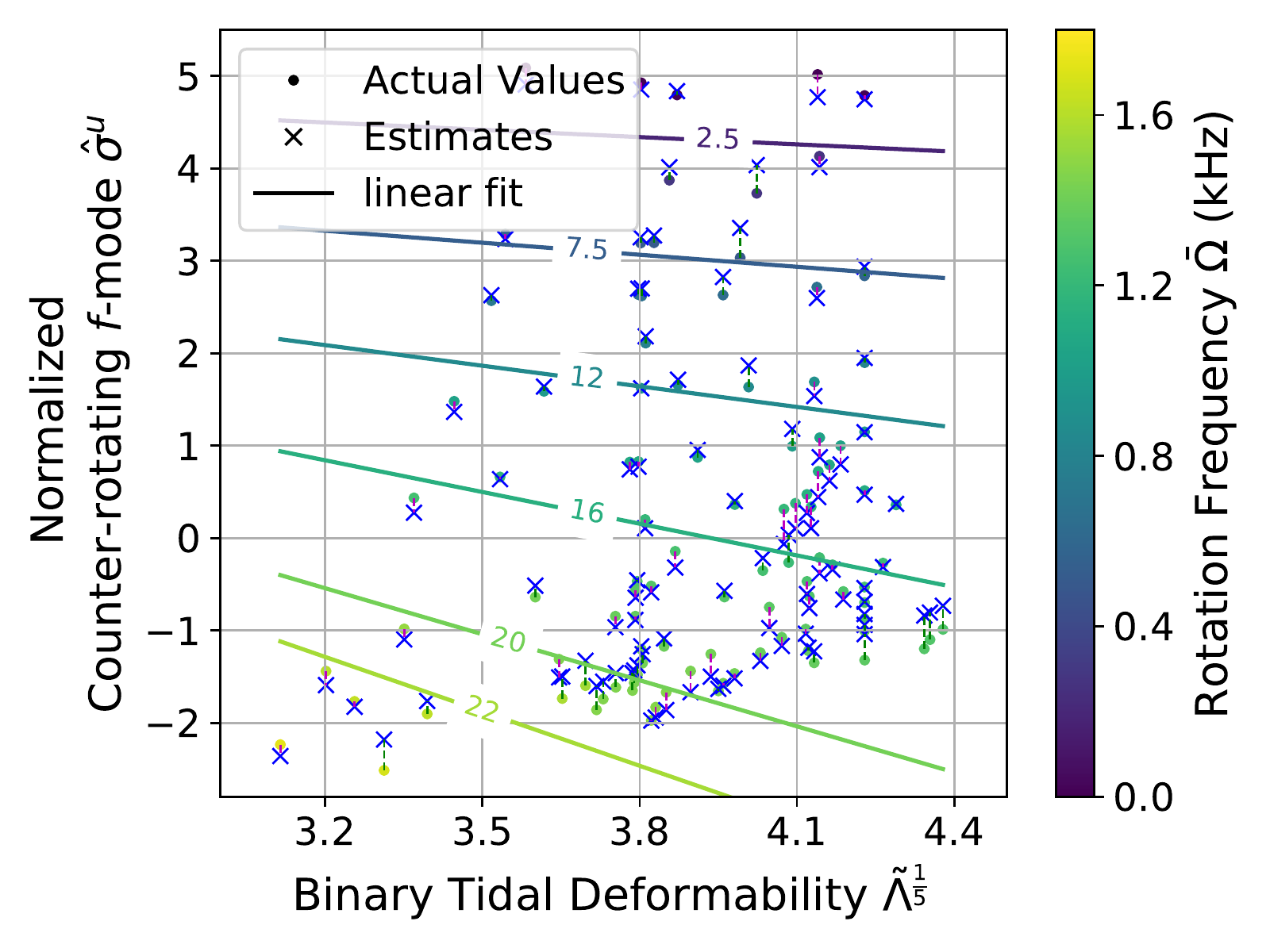}
\caption[Binary Tidal Deformability vs Counter-rotating $f$-mode Frequency for BNS with $q=1$.]{The counter-rotating $f$-mode frequency against the binary tidal deformability, for $q$=1. The linear fit corresponds to Equation~\eqref{eq:counter_fit}. The label on each line indicates the corresponding \emph{angular} rotation rate $\hat\Omega$. Note that negative frequencies correspond to the CFS-instability region of the counter-rotating mode.}
\label{fig:counter-f}
\end{figure}

In Figure~\ref{fig:counter-f}, we show the counter-rotating $f$-mode against the binary tidal deformability, colored by the rotation frequency of the star. We can immediately see that a single, rotation independent relation is clearly out of reach. However, we do observe a clear separation of the points by said rotation frequency, suggesting a linear relation of the form
\begin{equation}
\hat \sigma^u = a(q, \hat \Omega) \tilde \Lambda^\frac{1}{5} + b(q, \hat \Omega)
\label{eq:counter_fit}
\end{equation}
where 
\begin{equation}
\hat \Omega = \frac{M}{M_\odot}\frac{\Omega}{\si{k Hz}}
\label{eq:angular}
\end{equation}
is the \emph{normalized angular rotation rate}, and $a$ and $b$ are given by
\begin{equation}
a(q, \hat\Omega) = a_2(q) \hat\Omega^2 + a_1(q)\hat\Omega + a_0(q)
\label{eq:coeff-a}
\end{equation}
\begin{equation}
b(q, \hat\Omega) = b_2(q) \hat\Omega^2 + b_1(q)\hat\Omega + b_0(q)
\label{eq:coeff-b}
\end{equation}
where each of the $q$-dependent coefficients is again a quadratic function of $q$. Note that, while we did examine different parameterizations for the rotation rate, the normalized angular rotation rate $\hat\Omega$ proved to provide the best fit. 

Our best fit for this relation is also illustrated in Figure~\ref{fig:counter-f} for $\hat\Omega \in \{2.5, 7.5, 12, 16, 20, 22\}\, M_\odot \si{k Hz}$. We also show our estimates for all of the data points, each indicated by an `$\mathrm{x}$'. 
The lines connecting corresponding pairs are shown in pink if the estimate is smaller than the actual value, and in green if the estimate is larger.
\begin{table}[t]
\sisetup{round-mode=places,round-precision=3, scientific-notation=true}
\centering
\setlength\tabcolsep{2ex}
\renewcommand{\arraystretch}{1.4}
\begin{tabular}{c ? c | c | c }
$j$ & 2 & 1 & 0 \\\specialrule{.2em}{.1em}{.1em} 
$a_{2,j}$ & \num{0.00050323} & \num{-0.00098126} & \num{-0.00318426} \\
$a_{1,j}$ & \num{ 0.01083817} & \num{-0.02133844} & \num{0.01302439}   \\
$a_{0,j}$ & \num{0.05715282} & \num{-0.11308737} & \num{-0.19041603} \\ \specialrule{.2em}{.1em}{.1em} 
$b_{2,j}$ & \num{0.00056967} & \num{-0.00113636 } & \num{0.00798798}  \\
$b_{1,j}$ & \num{-0.05990596} & \num{0.11774872} & \num{-0.25715465} \\
$b_{0,j}$ & \num{-0.07978346} & \num{0.16054403} & \num{5.70890778} \\
\end{tabular}
\caption[Coefficients for Relation between Binary Tidal Deformability and Counter-rotating $f$-mode]{Coefficients of the relation for the counter-rotating $f$-mode in Equation~\eqref{eq:counter_fit}, considering only soft EoSs. The fit is illustrated in Figure~\ref{fig:counter-f}.}
\label{tab:coeff-counter}
\end{table}

The coefficients for this fit are listed Table~\ref{tab:coeff-counter}. They have the form
\begin{equation}
x_i = \sum_{j=0}^{2} x_{i,j} q^j
\label{eq:direct-coeff}
\end{equation}

This fit achieves an RMSE of $0.159$ within our data set, and an average relative error of $\sim 16\%$. Note that the relative error is necessarily higher here as we have frequency values around zero. 
If we only compare the RMSE with the error for the co-rotating relation, we achieve similar accuracy (note that there we estimated the logarithm of the normalized $f$-mode frequency). However, we now have a two-parameter fit, which is more difficult  to utilize in practice, as in addition to an estimate for the mass ratio, we would also need an accurate estimate for the rotation frequency of the remnant.

Still, in cases where we are able to obtain accurate estimates for these two quantities for a BNS merger, we can use the relation presented here to predict whether we might observe, e.g., a CFS-instability in the remnant due to a negative counter-rotating $f$-mode frequency~\cite{1970PhRvL..24..611C,1978ApJ...221..937F,2015PhRvD..92j4040D}.

\subsection{Combined \texorpdfstring{$\eta$}{n} Relation}\label{sec:combined}

Combining the linear $\tilde \Lambda - \hat\sigma^s$ fit in Equation~\eqref{eq:lin_fit} with the $\eta - \hat \sigma^s$ fit in~\cite[Equation (6)]{2020PhRvL.125k1106K}, we obtain a combined $\tilde \Lambda - \eta$ relation for the effective compactness of the form
\begin{equation}
\eta = \frac{10^{a(q) \cdot \tilde \Lambda^{\frac{1}{5}} + b(q)} - \left(c_1 + c_2 \hat \Omega + c_3 \hat \Omega^2\right)}{d_1 + d_3 \hat \Omega}
\label{eq:combined_relation}
\end{equation}
where $\hat \Omega$ is again the normalized angular rotation rate (cf. Equation~\eqref{eq:angular}) of the long-lived remnant, and the coefficients $c_1, c_2, c_3, d_1$ and $d_3$ are given in~\cite{2020PhRvL.125k1106K}. This relation allows us to estimate the effective compactness $\eta$ of the long-lived post-merger star from the tidal deformabilities of the pre-merger neutron stars.

\begin{figure}[t]
\includegraphics[width=0.5\textwidth]{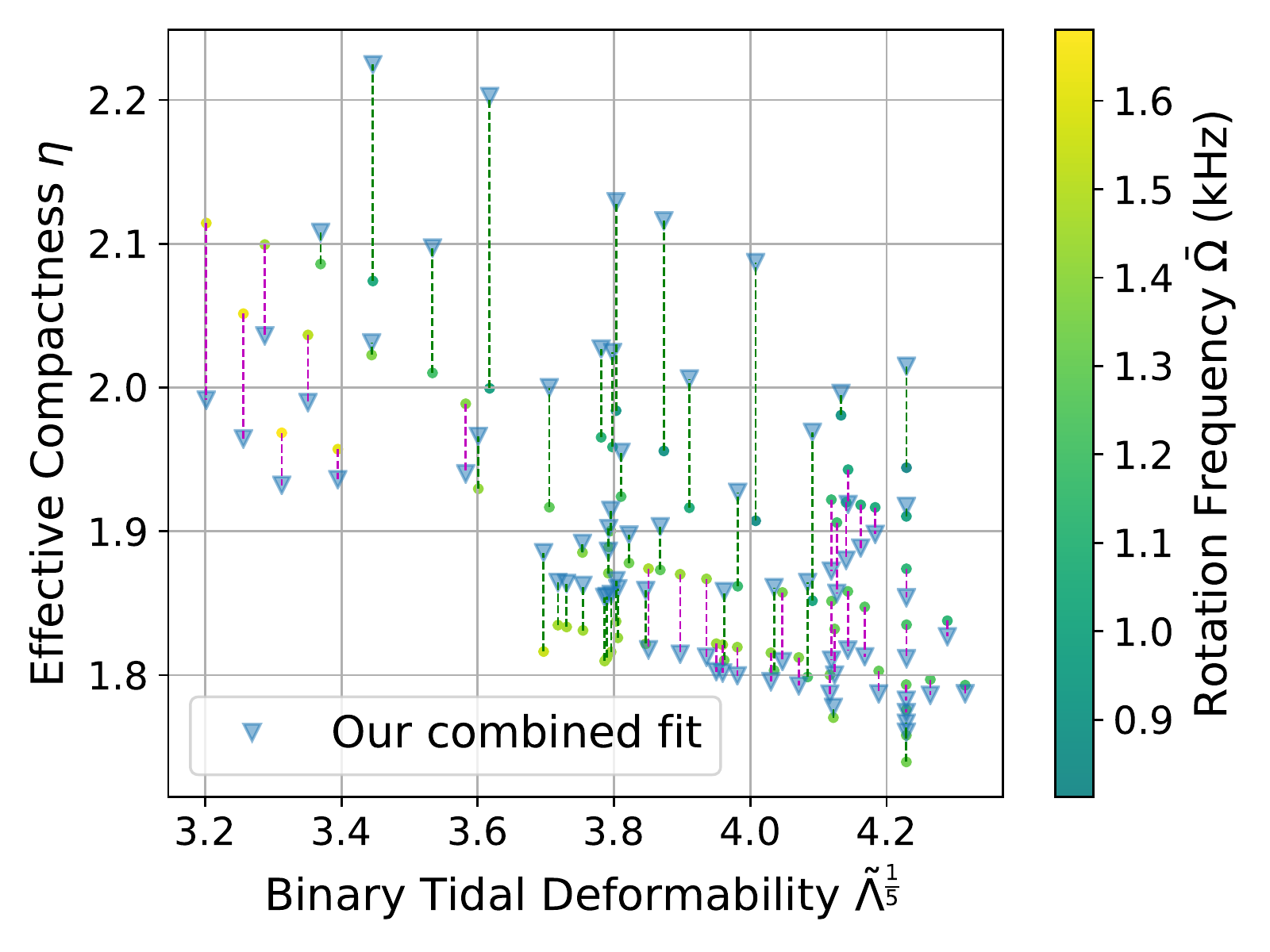}
\caption{Comparison of the estimated effective compactnesses through Equation~\eqref{eq:combined_relation} (triangles), and the actual effective compactnesses (dots) for $q=1$ and $\bar\Omega_{\mathrm{thr}} = 800 \si{Hz}$.}
\label{fig:combined}
\end{figure}

A comparison of the estimated effective compactnesses through Equation~\eqref{eq:combined_relation} (shown as triangles) and the actual effective compactnesses (shown as dots) is illustrated in Figure~\ref{fig:combined}. Again, lines connecting corresponding pairs are shown in pink if the estimate is smaller than the actual value, and in green if the estimate is larger.

The RMSE, average relative error and the maximum relative error for each rotation frequency threshold $\bar\Omega_{\mathrm{thr}}$ are given in Table~\ref{tab:combined_res}. On average, we achieve a relative error $\bar e$ of around $2\%-4\%$.
\begin{table}
\sisetup{round-mode=places,round-precision=3}
\centering
\setlength\tabcolsep{2ex}
\renewcommand{\arraystretch}{1.4}
\begin{tabular}{c ? c | c | c }
$\bar\Omega_{\mathrm{thr}} [\si{Hz}]$ & $400$ & $800$ & $1200$\\
\specialrule{.2em}{.1em}{.1em} 
$\overline{E}$ & \num{0.12223696411296842} & \num{0.06206404816178692} & \num{0.04259189290829366} \\
$\bar e$ & \num{0.03914872025548107}  & \num{0.024138501902225905} & \num{0.01929437772550319} \\
$e_{max}$ & \num{0.2155968218143253} & \num{0.10255421681860791} & \num{0.05386368611808994}\\
\end{tabular}
\caption[Errors of the Combined Fit for Different Rotation Frequency Thresholds]{The RMSE $\overline{E}$, the average relative error $\bar e$, and the maximum relative error $e_{max}$ of the combined fit (cf.~Equation~\eqref{eq:combined_relation}) at different rotation frequency thresholds $\bar\Omega_{\mathrm{thr}}$.}
\label{tab:combined_res}
\end{table}

Conceptually, we could formulate a similar relation using the counter-rotating $f$-mode, as well. However, the combined relation presented here is only intended as a motivating starting point, and will be replaced by a direct relation in the next section. We therefore do not further detail this approach at this point. 
\section{Direct Universal Relations}\label{sec:direct}
After using the results presented in Section~\ref{sec:f-mode} and in~\cite{2020PhRvL.125k1106K} to derive a combined relation between the pre-merger binary tidal deformability and the effective compactness of the long-lived remnant via the $f$-mode frequency, we now attempt a direct relation between these two quantities with the goal of achieving improved accuracy.
\subsection{Two-Parameter, Linear Relation}\label{sec:direct_linear}
\begin{figure}
    \includegraphics[width=0.5\textwidth]{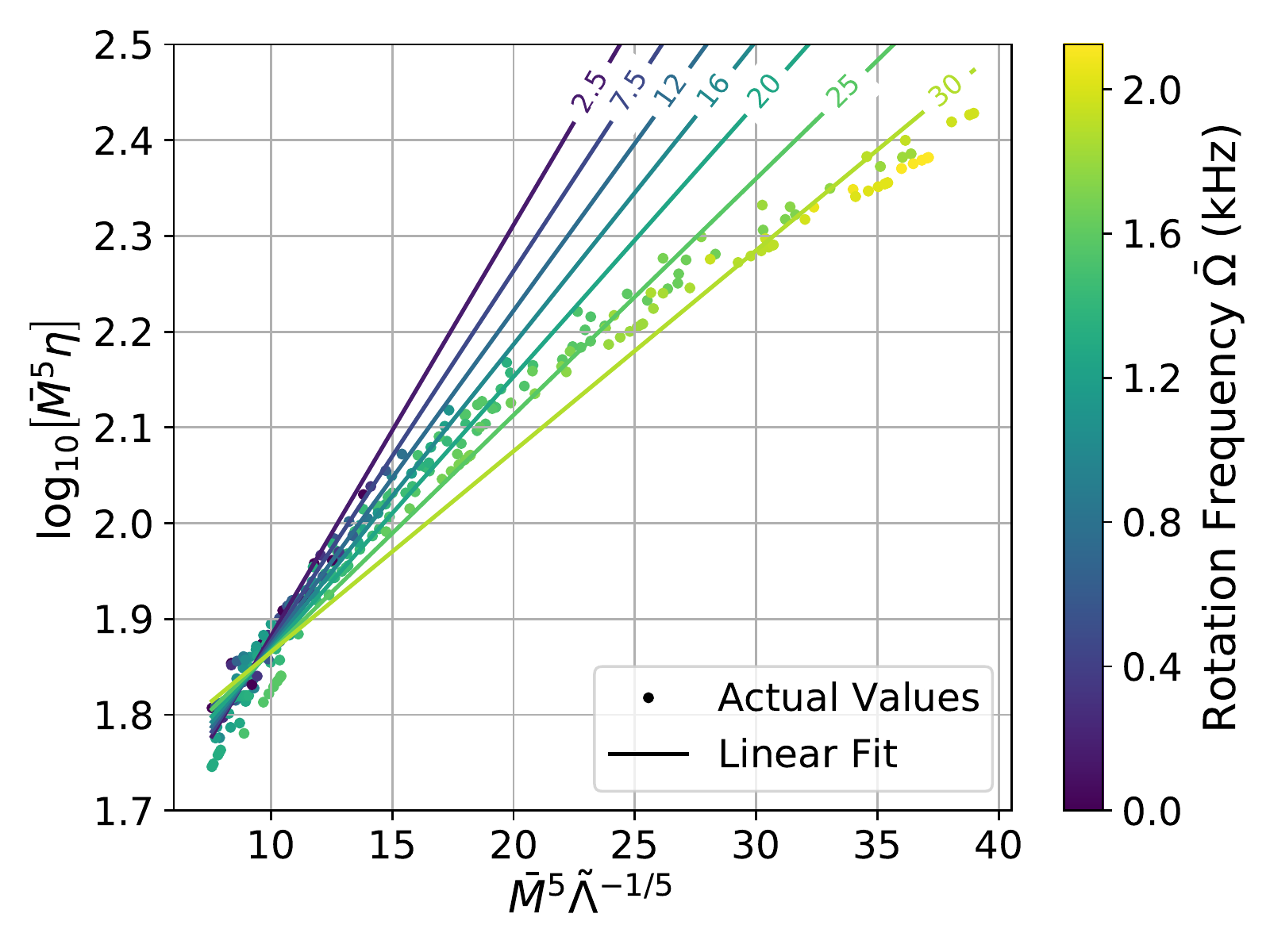}
    \caption{The relation between $\bar M^{5} \tilde \Lambda^{-\frac{1}{5}}$ and $\log_{10}\left[\bar M^5 \eta\right]$ for $q=1$, considering only soft EoSs. The linear, two-parameter fit corresponds to Equation~\eqref{eq:2-param-fit}. The labels on the lines indicate the corresponding \emph{angular} rotation rate $\hat \Omega$.}
    \label{fig:2-paramater_lambda_eta_fit}
\end{figure}

As with the combined relation in Section~\ref{sec:combined}, we begin by considering the EoSs that we call \emph{soft}, i.e. WFF1, APR4 and SLy (cf. Section~\ref{sec:equi}). 
A good fit is achieved by comparing $\bar M^{5} \tilde \Lambda^{-\frac{1}{5}}$ with $\bar M^5 \eta$. The resulting relation is illustrated in Figure~\ref{fig:2-paramater_lambda_eta_fit}.

We can observe a clear separation of the data points by the rotation frequency $\bar \Omega$. After examining different rotation parameters, the normalized angular rotation rate $\hat \Omega$ (cf. Equation~\eqref{eq:combined_relation}) again proved to perform best in our fits. Ultimately, we reach a linear, two-parameter fit of the form
\begin{equation}
\log_{10}\left[\bar M^5 \eta\right] = a(q, \hat\Omega) \bar M^{5} \tilde \Lambda^{-\frac{1}{5}} + b(q, \hat\Omega)
\label{eq:2-param-fit}
\end{equation}
with 
\begin{equation}
a(q, \hat\Omega) = a_2(q) \hat\Omega^2 + a_1(q)\hat\Omega + a_0(q)
\end{equation}
\begin{equation}
b(q, \hat\Omega) = b_2(q) \hat\Omega^2 + b_1(q)\hat\Omega + b_0(q)
\end{equation}
where each of the $q$-dependent coefficients is again a quadratic function of $q$.

The best fit of Equation~\eqref{eq:2-param-fit} to our data is illustrated in Figure~\ref{fig:2-paramater_lambda_eta_fit} 
for the case of $q = 1$. The errors incurred over all mass ratios are given in the first row of Table~\ref{tab:constraint_soft}: the fit achieves an RMSE of $\bar E \sim 0.020$, an average relative error of $\bar e \sim 0.7\%$, and a maximum relative error of $ e_{max} \sim 3.1\%$. The coefficients of this fit are listed in Table~\ref{tab:coeff-direct-soft}. They are defined as in Equation~\eqref{eq:direct-coeff}. The direct relation thus shows improved accuracy compared to the combined relation presented earlier: we go from an average relative error $\bar e$ of $2.4\%$ to $0.7\%$. 

\begin{table}[t]
\sisetup{round-mode=places,round-precision=3,scientific-notation=true}
\centering
\setlength\tabcolsep{2ex}
\renewcommand{\arraystretch}{1.2}
\begin{tabular}{c ? c | c | c }
$j$ & 2 & 1 & 0 \\\specialrule{.2em}{.1em}{.1em} 
$a_{2,j}$ & \num{-3.02998665e-06} & \num{5.92903281e-06} & \num{-4.44769773e-07} \\
$a_{1,j}$ & \num{ 0.00013551} & \num{-0.00026538} & \num{-0.00074878}   \\
$a_{0,j}$ & \num{0.00323056} & \num{-0.00631913} & \num{0.04821403} \\ \specialrule{.2em}{.1em}{.1em} 
$b_{2,j}$ & \num{-5.33135248e-05 } & \num{1.04739431e-04} & \num{-5.43100175e-05}  \\
$b_{1,j}$ & \num{-0.00057644} & \num{0.00112355} & \num{0.00692892} \\
$b_{0,j}$ & \num{0.01505976} & \num{-0.02946918} & \num{1.44849877} \\
\end{tabular}
\caption[Coefficients for Direct Relation between Binary Tidal Deformability and Effective Compactness]{Coefficients of the direct relation in Equation~\eqref{eq:2-param-fit}, considering only soft EoSs. The fit is illustrated in Figure~\ref{fig:2-paramater_lambda_eta_fit}.}
\label{tab:coeff-direct-soft}
\end{table}

\begin{figure}
    \includegraphics[width=0.5\textwidth]{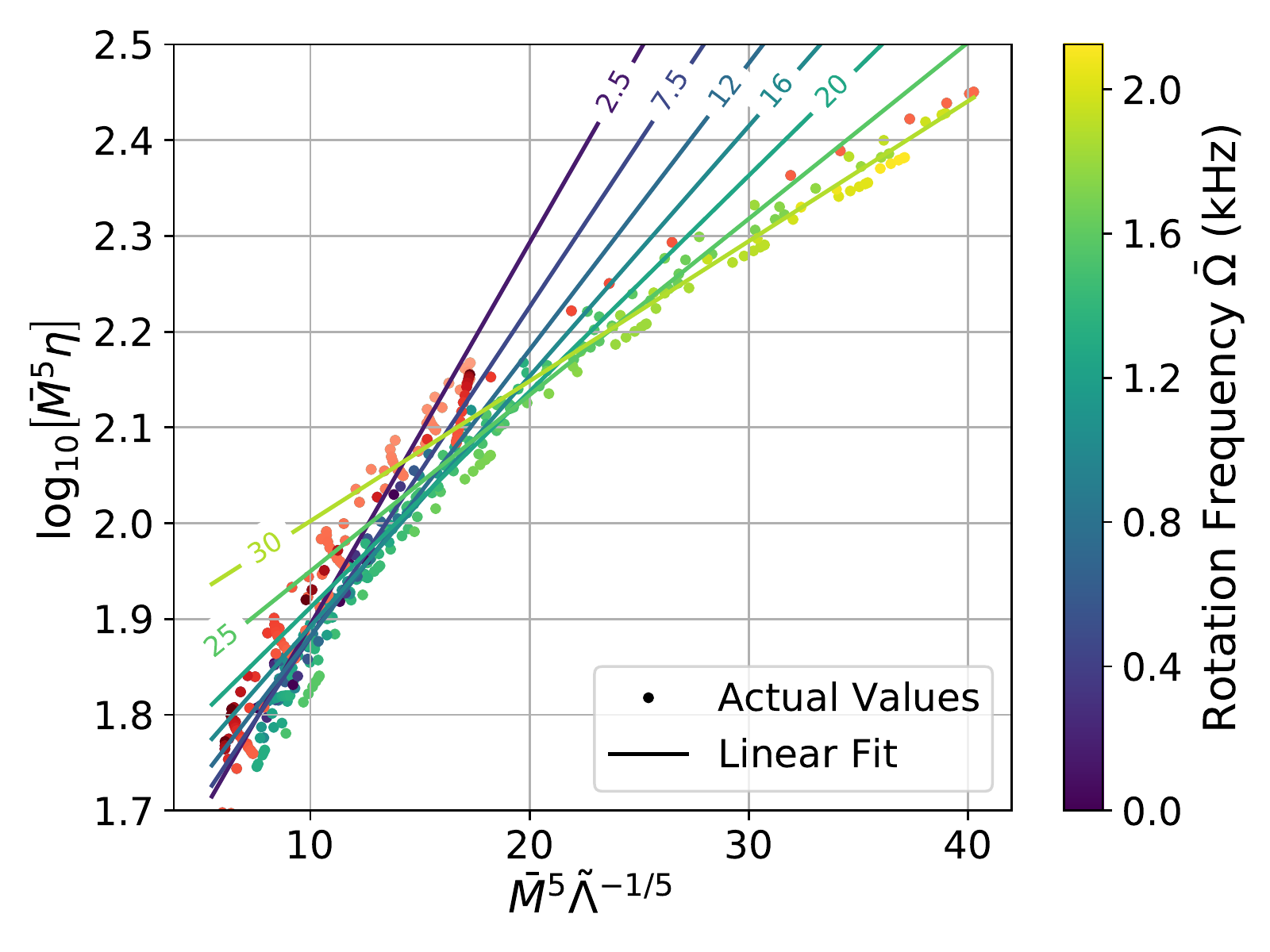}
    \caption{The same as in Figure~\ref{fig:2-paramater_lambda_eta_fit}, but with soft and stiff EoSs. Stiff EoSs are indicated in red.}
    \label{fig:2-paramater_lambda_eta_fit_all}
\end{figure}

With this improved accuracy, we can now also attempt to reintroduce the stiff EoSs (H4 and MS1) that we previously excluded for the combined relation.
The resulting best fit is illustrated in Figure~\ref{fig:2-paramater_lambda_eta_fit_all}. Here, the data points for the stiff EoSs are indicated in red.
This new fit for all EoSs achieves an RMSE of $\bar E \sim 0.045$, an average relative error of $\bar e \sim 1.7\%$, and a maximum relative error of $e_{max} \sim 7.3\%$. The coefficients of this fit can be found in Table~\ref{tab:coeff-direct-all}. They are again defined as in Equation~\eqref{eq:direct-coeff}.
\begin{table}
\sisetup{round-mode=places,round-precision=3,scientific-notation=true}
\centering
\setlength\tabcolsep{2ex}
\renewcommand{\arraystretch}{1.2}
\begin{tabular}{c ? c | c | c }
$j$ & 2 & 1 & 0 \\\specialrule{.2em}{.1em}{.1em} 
$a_{2,j}$ &  \num{-1.43299004e-06} & \num{2.79049501e-06} & \num{5.83975562e-06}\\
$a_{1,j}$ & \num{-8.12055734e-06} & \num{1.55508888e-05} & \num{-1.16100063e-03} \\
$a_{0,j}$ & \num{0.00316153} & \num{-0.00617711} & \num{0.04576627} \\\specialrule{.2em}{.1em}{.1em}
$b_{2,j}$ & \num{4.28711089e-05} & \num{-8.29514410e-05} & \num{2.58614274e-04} \\
$b_{1,j}$ & \num{-0.00120265} & \num{0.0023401} & \num{0.00489851} \\
$b_{0,j}$ &  \num{0.01130244} & \num{-0.022166} & \num{1.48913597} \\
\end{tabular}
\caption[Coefficients for Direct Relation between Binary Tidal Deformability and Effective Compactness, with stiff EoSs]{Coefficients of the direct relation in Equation~\eqref{eq:2-param-fit}, considering soft and stiff EoSs. The fit is illustrated in Figure~\ref{fig:2-paramater_lambda_eta_fit_all}.}
\label{tab:coeff-direct-all}
\end{table}

As we can see, the direct relation admits the stiff EoSs with close to double the error. However, the error remains small and we achieve high accuracy even including stiff EoSs, thus improving the universality of our direct relation. 

The inclusion of the stiff EoSs has a similar effect on the other relations we present below. For the sake of brevity, we will therefore present the results including only the soft EoSs in the remainder of the text, and refer to the appendix for the results with all EoSs. 

\begin{table}
\sisetup{round-mode=places,round-precision=3}
\centering
\renewcommand{\arraystretch}{1.4}
\setlength\tabcolsep{1.5ex}
\begin{tabular}{c ? c | c | c}
& $\overline{E}$ &$\bar e$ & $e_{max}$ \\\specialrule{.2em}{.1em}{.1em} 
linear, two-parameter & \num{0.019561860286427393} & \num{0.007492819154409199} &  \num{0.031094088675910347} \\
quadratic, one-parameter & \num{0.02109185291028902} & \num{0.008562377582848873} & \num{0.03119307374870066} \\ 
constrained $\Omega$ & \num{0.014479835735132687} & \num{0.006013420468363995} & \num{0.018039020406442507} \\
\end{tabular}
\caption[Errors of Direct Relation between Binary Tidal Deformability and Effective Compactness]{RMSE $\bar E$, average relative error $\bar e$ and maximum relative error $e_{max}$ achieved for the relations in Equations~\eqref{eq:2-param-fit} (linear, two-param.) and \eqref{eq:quadratic-fit} (quadratic, one-param.), only considering soft EoSs. The constrained relation is introduced in Section~\ref{sec:constrained}.}
\label{tab:constraint_soft}
\end{table}

\subsection{One-Parameter, Quadratic Relation}\label{sec:direct_quadratic}
\begin{figure}
    \includegraphics[width=0.5\textwidth]{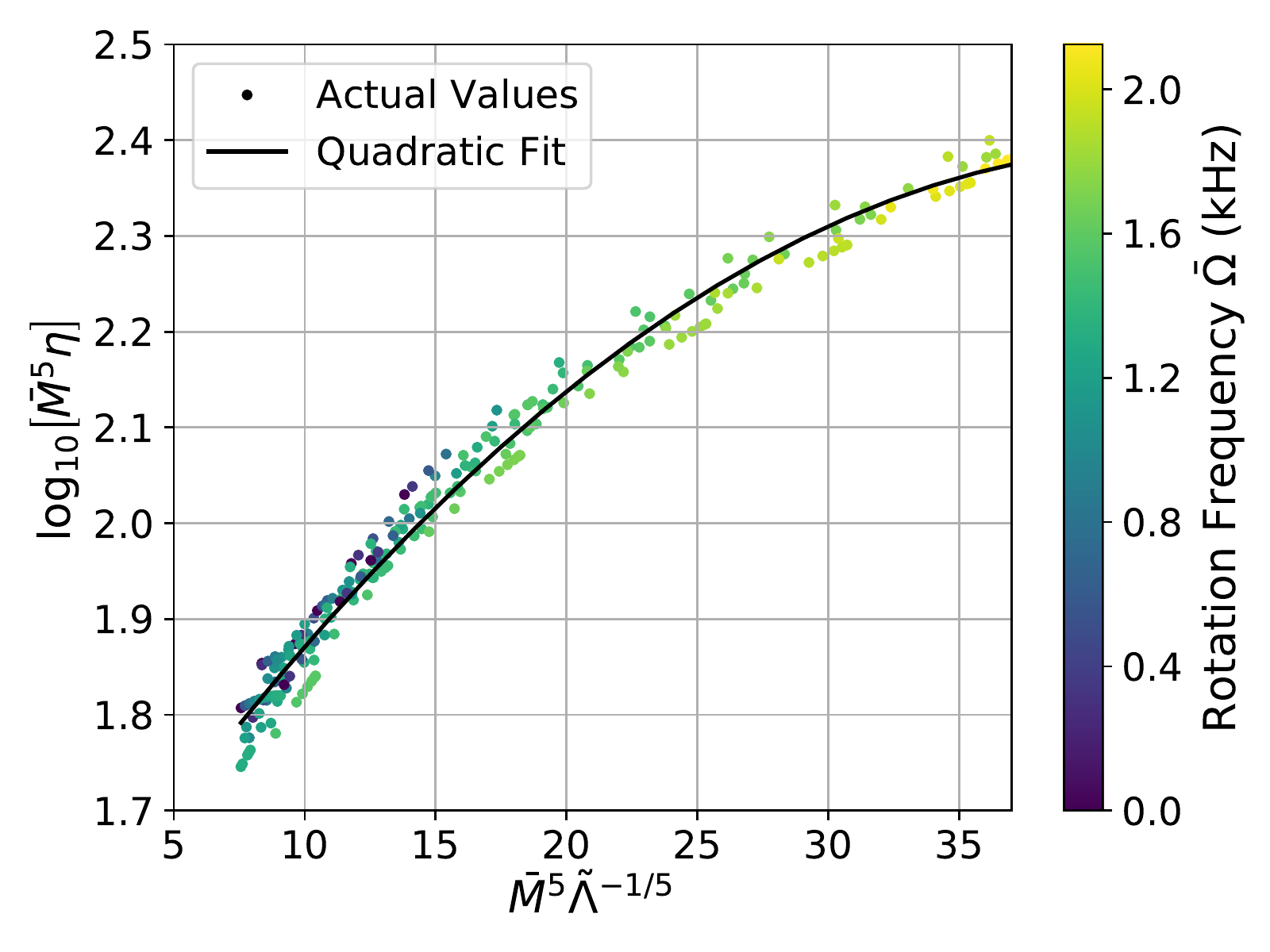}
    \caption{The quadratic fit between $\bar M^{5} \tilde \Lambda^{-\frac{1}{5}}$ and $\log_{10}\left[\bar M^5 \eta\right]$ for $q=1$, considering only soft EoSs. The fit corresponds to Equation~\eqref{eq:quadratic-fit}.}
    \label{fig:quadratic_lambda_eta_fit_soft}
\end{figure}

The plots in Figures~\ref{fig:2-paramater_lambda_eta_fit} and~\ref{fig:2-paramater_lambda_eta_fit_all} suggest that a one-parameter fit that only depends on the mass ratio $q$, and is independent of the rotation frequency, might also be possible. We therefore attempt a relation of the form
\begin{equation}
\log_{10}\left[\bar M^5 \eta\right] = a(q) \left(\bar M^{5} \tilde \Lambda^{-\frac{1}{5}}\right)^2 + b(q) \bar M^{5} \tilde \Lambda^{-\frac{1}{5}} + c(q)
\label{eq:quadratic-fit}
\end{equation}
where each of the coefficients $a, b$ and $c$ are quadratic in $q$ and take the form 
\begin{equation}
x = \sum_{i=0}^{2} x_{i} q^i.
\label{eq:coeff}
\end{equation}
The best fit of our data to this relation is shown in Figure~\ref{fig:quadratic_lambda_eta_fit_soft}. The coefficients corresponding to this fit are listed in Table~\ref{tab:coeff-quadratic_soft} and the resulting errors in the second row of Table~\ref{tab:constraint_soft}. 

The quadratic fit performs only slightly worse than the linear, two-parameter fit, going from an average relative error of $0.8\%$ for the two-parameter fit, to $0.9\%$ for the one-parameter fit. 
However, it only depends on one free parameter, the mass ratio $q$, and no longer on the rotation rate of the remnant. As such, the quadratic fit is preferable to the linear two-parameter fit in practice as we reduce the number of parameters
that need to be estimated, while still achieving very similar accuracy. 

\begin{table}[t!]
\sisetup{round-mode=places,round-precision=3,scientific-notation=true}
\centering
\setlength\tabcolsep{1ex}
\renewcommand{\arraystretch}{1.4}
\begin{tabular}{ c ? c | c | c }
$i$ & 2 & 1 & 0 \\\specialrule{.2em}{.1em}{.1em}
$a_i$ & \num{-0.00012387} & \num{ 0.0002423} & \num{-0.00058986} \\
$b_i$ & \num{0.00558613} & \num{-0.01093207} & \num{0.04616257} \\
$c_i$ & \num{-0.00617142} & \num{0.01210978} & \num{1.50405505}
\end{tabular}
\caption{Coefficients of the quadratic, one-parameter relation in Equation~\eqref{eq:quadratic-fit}, only considering soft EoSs. The fit is illustrated in Figure~\ref{fig:quadratic_lambda_eta_fit_soft}.}
\label{tab:coeff-quadratic_soft}
\end{table}

\subsection{Compactness Relation}\label{sec:compactness}
The compactness of neutron star has been the object of various universal relations considered in the past~\cite{1998MNRAS.299.1059A,2004PhRvD..70l4015B,PhysRevLett.95.151101,PhysRevD.91.044034}. The advantage of an accurate universal relation that connects the binary tidal deformability of the BNS to the compactness of the long-lived remnant is that it allow us to directly constrain the remnant's radius (using independent estimates for its gravitational mass), and through that its EoS. We therefore now investigate the possibility of such a relation. 

We compute the compactness of the long-lived remnant through $C = M/R_e$ where $R_e$ is its equatorial radius.
\begin{figure}
    \includegraphics[width=0.5\textwidth]{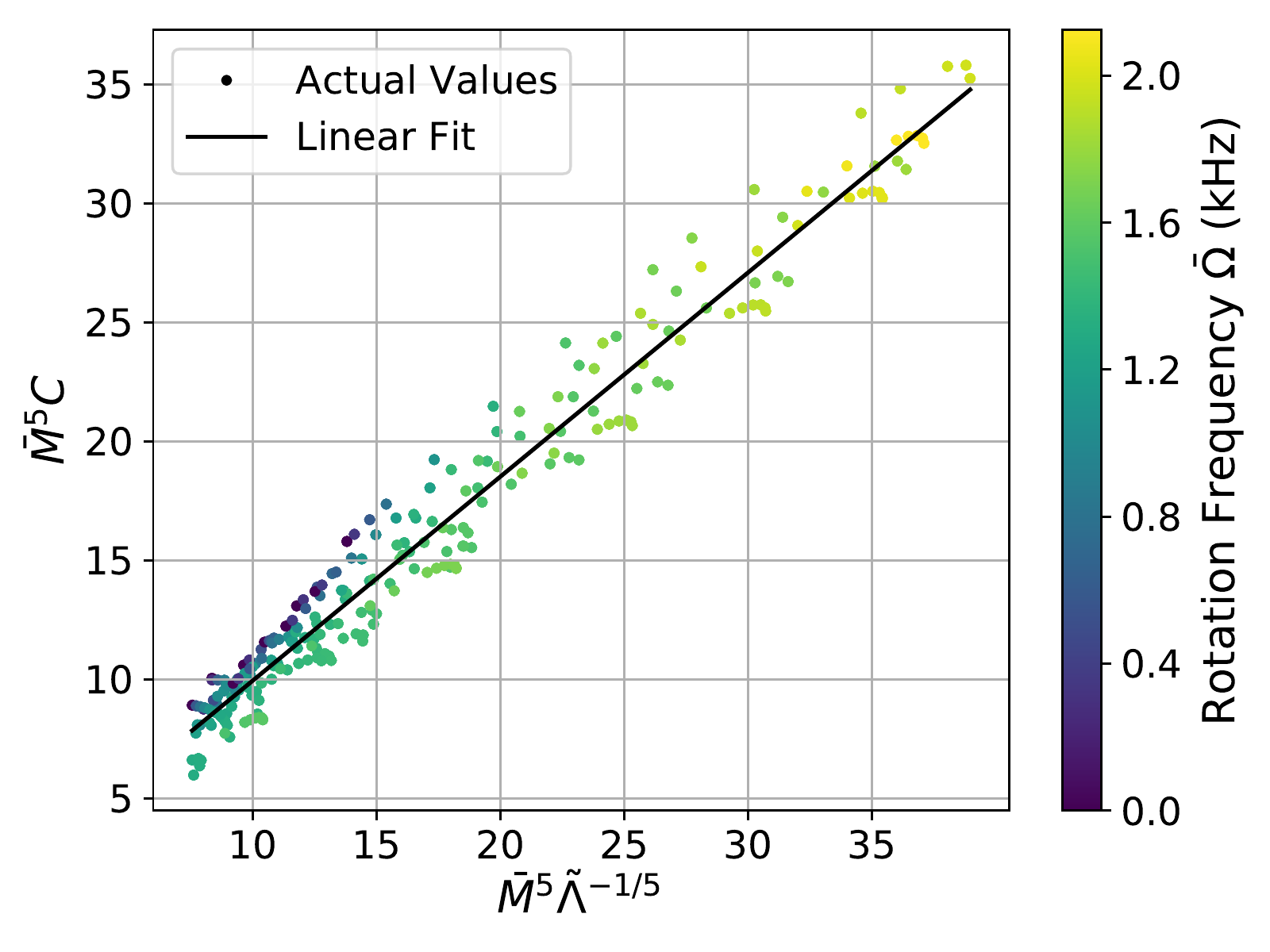}
    \caption{The linear fit between $\bar M^{5} \tilde \Lambda^{-\frac{1}{5}}$ and $\bar M^5 C$ for $q=1$, considering only soft EoSs. The fit corresponds to Equation~\eqref{eq:compactness-fit}.}
    \label{fig:compactness_lambda_fit_soft}
\end{figure}
After considering several options, we find that the best fit is obtained for a relation of the form
\begin{equation}
\bar M^5 C = a(q) \bar M^{5} \tilde \Lambda^{-\frac{1}{5}} + b(q)
\label{eq:compactness-fit}
\end{equation}
where the coefficients $a$ and $b$ have the same (quadratic) form as given in Equation~\eqref{eq:coeff}.

The best fit for Equation~\eqref{eq:compactness-fit} is illustrated in Figure~\ref{fig:compactness_lambda_fit_soft}. The coefficients of the best fit are given in Table~\ref{tab:coeff-compactness_soft}.
\begin{table}
\sisetup{round-mode=places,round-precision=3}
\centering
\setlength\tabcolsep{1ex}
\renewcommand{\arraystretch}{1.4}
\begin{tabular}{c ? c | c | c }
 $i$ & 2 & 1 & 0 \\\specialrule{.2em}{.1em}{.1em}
$a_i$ & \num{0.13054156} & \num{-0.25547174} & \num{0.98258018} \\
$b_i$ & \num{-0.32197869} & \num{0.63141607} & \num{1.04875516} \\
\end{tabular}
\caption{Coefficients of the linear, one-parameter relation in Equation~\eqref{eq:compactness-fit}, considering only soft EoSs. The fit is illustrated in Figure~\ref{fig:compactness_lambda_fit_soft}.}
\label{tab:coeff-compactness_soft}
\end{table}
As we can see, the spread of the data points for the compactness relation is higher than for the effective compactness relations considered above. As listed in the first row of Table~\ref{tab:compactness_errors_soft},
the compactness relation achieves an RMSE of $1.326$ and an average relative error of $7.6\%$. These errors are around an order of magnitude larger than for the effective compactness relations. Thus, the compactness relation clearly shows worse universality. 

Despite the notably worse accuracy of the compactness relation, it might still be useful in practice by allowing us to directly constrain the remnant's (equatorial) radius. Of course, the worse accuracy of this relation would also impact the accuracy of the radius constraints obtained in such a way.

\section{Relations with Constrained Rotation Rates}\label{sec:constrained}
The universal relations discussed above allows for the full range of rotation frequencies of the long-lived remnant, from non-rotating up to the Keplerian limit.
As such, imposing additional constraints to the remnant's rotation rate appears to be a straightforward way to further improve the accuracy of our relations. 

As we discussed in Section~\ref{sec:methods}, Radice et al.~\cite{2018MNRAS.481.3670R} propose a universal relation for the spin-period of long-lived remnants (cf. Equation~\eqref{eq:rot_constraint}) of BNS mergers, which they derived from numerical
relativity simulations. We apply this universal relations here to our own long-lived remnants to constrain their rotation rates: given the baryon mass of the remnant, we only allow rotation rates that are predicted by Equation~\eqref{eq:rot_constraint}.

Through this, we end up with rotation frequencies around $\bar \Omega \sim 950-1450 \si{Hz}$ (depending on the remnant's baryon mass). 
For our evaluations, we only consider neutron stars with rotation rates in a $10\%$ range around this value. 
In the following, we investigate how much the accuracy of the one-parameter relations for the effective compactness (cf. Section~\ref{sec:direct_quadratic})  and the compactness (cf. Section~\ref{sec:compactness}) relations improves by introducing this rotation rate constraint. 
We list the resulting errors for the effective compactness relation in the third row of Table~\ref{tab:constraint_soft}, and for the compactness relation in the second row of Table~\ref{tab:compactness_errors_soft}. 

Overall, this constraint only has a small impact on the accuracy of our relations: for the effective compactness relation, for instance, we go from an average relative error of $0.9\%$ from the quadratic relation without constraint (cf. second row of Table~\ref{tab:constraint_soft}), to $0.6\%$ with constraint. We observe the largest change with the maximum relative error: it improves from $3.1\%$ without constraint, to $1.8\%$ with constraint.

The effect is similar for the compactness relation: the relative error again shows a slight improvement, from $7.6\%$ without constraint, to $6.1\%$ with constraint. The maximum relative error again shows a more significant improvement, going from $32.2\%$ for the unconstrained case, to $17.8\%$ in the constrained case, mirroring our observations for the effective compactness relation.

As such, we have found that introducing constraints on the free parameters in our relations can indeed improve the accuracy, albeit only slightly. Investigating how adding further constraints could more significantly improve our relations 
could be a fruitful direction for future work.

\begin{table}
\sisetup{round-mode=places,round-precision=3}
\centering
\renewcommand{\arraystretch}{1.4}
\setlength\tabcolsep{1.5ex}
\begin{tabular}{c ? c | c | c}
& $\overline{E}$ &$\bar e$ & $e_{max}$ \\\specialrule{.2em}{.1em}{.1em} 
unconstrained & \num{1.3259756483516871} & \num{0.07590372537615378} &  \num{0.32199587278912795} \\
constrained & \num{1.1444383062794299} & \num{0.06054441695002255} & \num{0.17848304482150495} \\ 
\end{tabular}
\caption{RMSE $\bar E$, average relative error $\bar e$ and maximum relative error $e_{max}$ achieved for the linear compactness relation (cf. Equation~\eqref{eq:compactness-fit}), with and without constraint on the remnant's rotation rate, only considering soft EoSs.}
\label{tab:compactness_errors_soft}
\end{table}

\section{Mass Loss During Spin Down}\label{app:mass_loss}
Until now, we have assumed a negligible baryon mass loss going from the pre-merger neutron stars to the long-lived remnant of the merger. As we discussed in Section~\ref{sec:methods}, while the baryon mass loss during the merger is indeed negligible~\cite{2013ApJ...773...78B,2013PhRvD..87b4001H,2016PhRvD..93l4046S,2018ApJ...869..130R}, Radice et al.~\cite{2018MNRAS.481.3670R} put an upper bound of $0.2 M_\odot$ baryon mass lost during the spin down of the early, differentially rotating remnant to the long-lived, uniformly rotating remnant. Since this potential mass loss is substantial, we here revisit our universal relations for the effective compactness under the assumption that a baryon mass of $\Delta M \in \{0.1 M_\odot, 0.2 M_\odot\}$ is lost during the transition from the pre-merger stars to the long-lived remnant, i.e.
\begin{equation}
M_b = M_{b,1} + M_{b,2} - \Delta M
\end{equation} 
where $M_b$ is the baryon mass of the long-lived remnant, and $M_{b,1}$ and $M_{b,2}$ the baryon masses of the pre-merger stars.
\begin{figure*}
\subfloat[$\Delta M = 0.1 M_\odot$]{\includegraphics[width=0.5\textwidth]{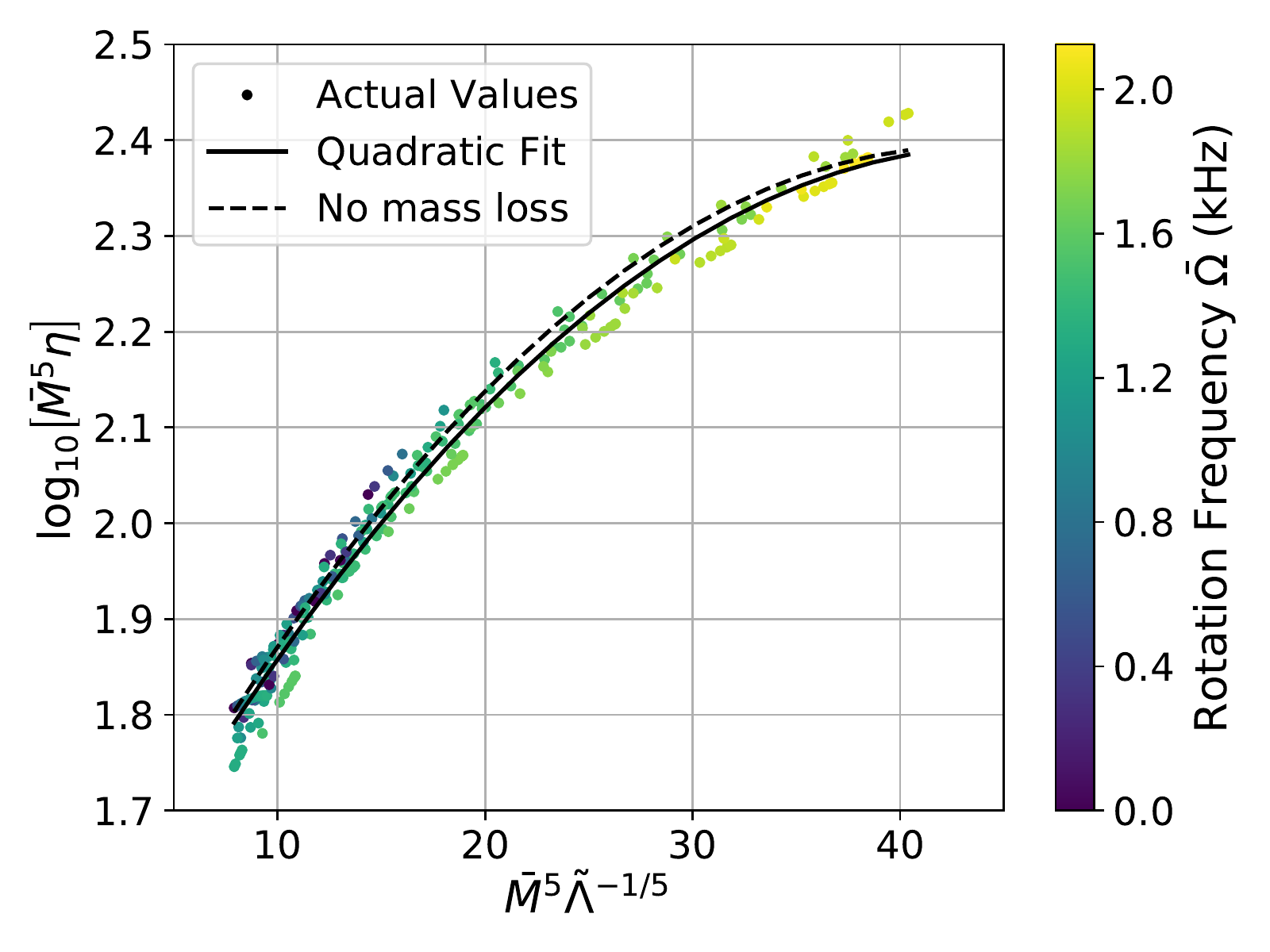}}
\subfloat[$\Delta M = 0.2 M_\odot$]{\includegraphics[width=0.5\textwidth]{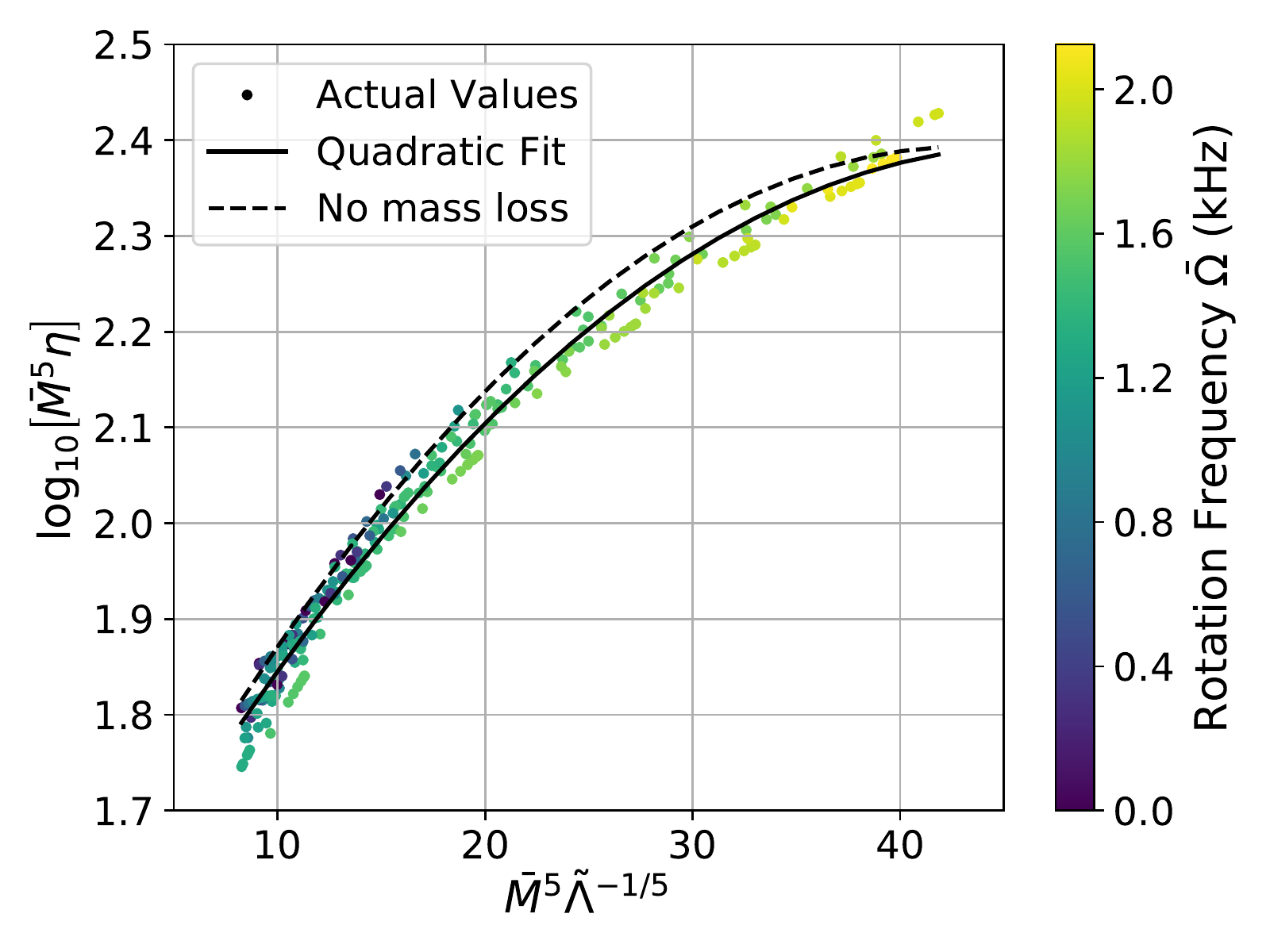}}
\caption{The quadratic fit (cf. Equation~\eqref{eq:quadratic-fit}) between $\bar M^{5} \tilde \Lambda^{-\frac{1}{5}}$ and $\log_{10}\left[\bar M^5 \eta\right]$ for $q=1$, with a $\Delta M$ baryon mass loss, considering only soft EoSs. The dashed line represents the original fit.}
\label{fig:app-01_mass_loss-quad}
\end{figure*}

Since we have a fixed data set of remnant configurations from~\cite{2020PhRvL.125k1106K}, we implement the mass loss by keeping the remnant configuration fixed, and instead increasing the masses of the pre-merger stars to account for the $\Delta M$ mass loss. We then recompute the solution of the TOV equations and the tidal deformability for this new set of pre-merger stars. In general, the inclusion of a non-negligible baryon mass loss will therefore cause a slight shift on the $\bar M^5 \tilde \Lambda^{-\frac{1}{5}}$ axis in Figures~\ref{fig:2-paramater_lambda_eta_fit} to~\ref{fig:compactness_lambda_fit_soft} as the increase of the pre-merger baryon masses will be reflected in a decrease of the binary deformability $\tilde \Lambda$.

Since the effect of the mass loss is essentially the same for all of the universal relations presented in this paper, we here exemplarily only discuss the quadratic, one-parameter relation introduced in Section~\ref{sec:direct_quadratic}. We here also also only discuss the case of soft EoSs, and refer to Appendix~\ref{app:mass_loss_all} for the case with all EoSs. However, the effect of including stiff EoSs is essentially the same as in the relations considered previously, and does not have any additional effect when combined with a non-negligible mass loss.

In the following, we first re-compute the new best fit of this relation to the new set of BNS merger data that assumes a baryon mass loss of $\Delta M$, and present the resulting errors. Afterwards, we try to apply our original relation (without mass loss) to the mergers with mass loss and give an estimate of the error incurred if the original relation were used for such mergers. 	

\subsection{New Best Fits for Mergers with Mass Loss}
\begin{table}
\sisetup{round-mode=places,round-precision=5}
\centering
\renewcommand{\arraystretch}{1.4}
\setlength\tabcolsep{1.5ex}
\begin{tabular}{c ? c | c | c}
$\Delta M$ & $\overline{E}$ &$\bar e$ & $e_{max}$ \\\specialrule{.2em}{.1em}{.1em} 
0 & \num{0.02109185291028902} & \num{0.008562377582848873} & \num{0.03119307374870066} \\ 
$0.1 M_\odot$ & \num{0.021039447677730742} & \num{0.008535745950222532} & \num{0.031060998161678118} \\ 
$0.2 M_\odot$ & \num{0.020988539280412083} & \num{0.008508802682963303} & \num{0.03092123630656561} \\ 
\end{tabular}
\caption[Errors of Direct Relation between Binary Tidal Deformability and Effective Compactness]{RMSE $\bar E$, average relative error $\bar e$ and maximum relative error $e_{max}$ achieved for the quadratic, one-parameter relation (cf. Equation~\eqref{eq:quadratic-fit}), with and without mass loss. Only considering soft EoSs.}
\label{tab:errors_mass_loss_comparison_quad}
\end{table}

The data points and new best fits of the quadratic, one-parameter relation for BNS mergers with $\Delta M = 0.1 M_\odot$ and $0.2 M_\odot$ are shown in Figure~\ref{fig:app-01_mass_loss-quad}. For the sake of comparison, we also show our original fit (without mass loss) as a dashed line. 

As we can see, the inclusion of mass loss does not change the form of our relation, or invalidate their universality. While there is a shift of the data points caused by the different pre-merger masses, the same relation as previously can be constructed 
with very similar errors. We list the errors of the new best fits together with the errors of our original relation in Table~\ref{tab:errors_mass_loss_comparison_quad}. We increased the number of decimal places shown compared to previous tables to show when the first differences appear. Ultimately, there is effectively no change in the accuracy of the best fits with increasing mass loss. 

\begin{table}[t]
\sisetup{round-mode=places,round-precision=3}
\centering
\renewcommand{\arraystretch}{1.4}
\setlength\tabcolsep{1.5ex}
\begin{tabular}{ c ? c | c | c}
 Mass Loss & $\overline{E}$ &$\bar e$ & $e_{max}$ \\\specialrule{.2em}{.1em}{.1em} 
 0 & \num{0.02109185291028902} &\num{0.008562377582848873} & \num{0.03119307374870066} \\
$0.1 M_\odot$ & \num{0.02541696569791403} & \num{0.009962398813830601} & \num{0.038071092679324896} \\ 
$0.2 M_\odot$& \num{0.03521206216595593} &\num{0.014801336317175234} & \num{0.04494964297558875}
\end{tabular}
\caption{RMSE $\bar E$, average relative error $\bar e$ and maximum relative error $e_{max}$ achieved by our original quadratic, one-parameter fit (cf. Equations~\eqref{eq:quadratic-fit}), for mergers with $0$, $0.1 M_\odot$ and $0.2 M_\odot$ baryon mass loss. Only considering soft EoSs.}
\label{tab:errors_02_mass_loss_quadratic}
\end{table}

\subsection{Original Relation for Mergers with Mass Loss}
We have seen above that the inclusion of a non-negligible baryon mass loss still allows us to construct universal relations between the pre-merger binary tidal deformability and the effective compactness of the long-lived remnant (assuming a good estimate for the actual baryon mass loss is known). We now estimate the error incurred by using the original relation (without mass loss) for the mergers with $0.1 M_\odot$ and $0.2 M_\odot$ baryon mass loss, respectively. We give the resulting errors in Table~\ref{tab:errors_02_mass_loss_quadratic}. 

As one would expect, the accuracy of our original relation (without mass loss) decreases as the mass loss $\Delta M$ increases. However, even for the worst case of $\Delta M = 0.2 M_\odot$, our original relation remains fairly accurate: 
we achieve an average relative error of $1.5\%$ if we only consider soft EoSs, compared to $0.9\%$ for mergers without mass loss.

Consequently, the relations presented in the paper seem to be robust against limited mass losses during the post-merger phase at the range of $0.2 M_\odot$ as predicted by Radice et al.~\cite{2018MNRAS.481.3670R}. Therefore, even if we are not able to further constrain the post-merger mass loss, our original relation will remain useful.

\section{Discussion \& Conclusion}\label{sec:conclusion}
In this paper, we proposed a novel approach to developing universal relations for BNS mergers using results from perturbative calculations
instead of using full-fledged numerical relativity simulations. 

Our approach is based on a simplified model for a BNS merger that results in a long-lived neutron star remnant: instead of performing a detailed 
simulation of the merger process, we individually consider the pre-merger phase, represented by two irrotational neutron stars, and the long-lived remnant,
represented by a rapidly rotating neutron star. By comparing the neutron stars in these two phases, we then obtain relations connecting their properties.

Inspired by previous results
by Kiuchi et al.~\cite{2020PhRvD.101h4006K}, and using the results from~\cite{2020PhRvL.125k1106K}, we proposed a novel universal 
relation between the pre-merger binary tidal deformability $\tilde\Lambda$ and the effective compactness $\eta$ of the long-lived 
remnant with high accuracy, reaching average relative error of $0.9\%$ if we only conisder soft EoSs, and $\sim 1.5\%$ if we also include stiff EoSs.

Similarly, we find a novel universal relation between the pre-merger binary tidal deformability and the compactness of the long-lived remnants, however with larger average relative error of 
$\sim 7.6\%$ if we only consider soft EoSs (and $8.8\%$ with all EoSs).

We furthermore find that the impact of a non-negligible mass loss during the transition from the differentially rotating, early remnant to the uniformly, rotating long-lived remnant on our relations is small. In particular, we find that a) we can easily construct new universal relations if good estimates for the post-merger baryon mass loss is known, and b) even the original relations (without mass loss) we presented in this paper can be applied to mergers with non-negligible baryon mass loss (up to $0.2 M_\odot$) with only a slight increase in the average relative error.

As the binary tidal deformability can be constrained fairly well from the observation of the gravitational waves emitted during the pre-merger 
phase~\cite{2008PhRvD..77b1502F,2014PhRvL.112j1101F}, and as the spectrum of the pre-merger gravitational waves lie very well within the sensitivity range of current-generation
GW detectors~\cite{2014ApJ...784..119R}, the relations presented in this paper can be useful for predicting the stellar parameters of a potentially long-lived remnant produced by a BNS merger. 

They can, in particular, be useful for further constraining the true EoS of neutron stars: for instance, as recently discussed by Greif et al.~\cite{2020ApJ...901..155G}, combining 
accurate estimates for the gravitational mass $M$ (found directly from the gravitational waves) and the moment of inertia $I$ (derived by using Equation~\eqref{eq:direct} to estimate $\eta = \sqrt{M^3/I}$), 
can put further constraints on the radius of the neutron star, and thus its EoS.

While we only used perturbative calculations for our analyses, they were not performed in a vacuum: numerical relativity simulations still play an important role in directing our efforts into the right direction. Not only were our efforts inspired by previous, simulation-based results by Kiuchi et al.~\cite{2020PhRvD.101h4006K}, but also, by employing the universal relation for the remnant spin period put forward by Radice et al.~\cite{2018MNRAS.481.3670R}, we were able to further constrain our universal relations to more physical rotation rates of the remnant, improving our accuracy, and reducing the number of free parameters. As such, the approach presented in this paper should mostly be considered supplementary to existing numerical relativity efforts. 

\medskip
\noindent
\textbf{Future Directions.}
As already mentioned in the introduction, our model for the BNS merger is markedly simple and does not cover all
aspects and details relevant for the merger of two neutron stars. Our approach, however, can be freely extended 
to include more involved measures such as hot EoSs to model the remnant right after the merger before it cools down.
As discussed in Section~\ref{sec:methods}, considering EoSs with phase transitions, specifically, will be an important direction for future work.

Of interest would also be to include differential rotation into our post-merger model and to compute the corresponding
oscillation modes (and other stellar parameters). This task has already been performed for simple differential rotation models under various approximations, 
such as the Cowling approximation~\cite{2002ApJ...568L..41Y,PhysRevD.81.084019}. However, a more comprehensive approach
with a more accurate model of differential rotation, and the treatment of the oscillations modes without approximation, would be desirable.

\begin{acknowledgments}
C.K. gratefully acknowledges financial support by DFG research Grant No. 413873357.
\end{acknowledgments}

\appendix
\section{Equilibrium Model}\label{app:equi}
\begin{figure*}[t]
\subfloat[$\tilde\Lambda - \hat \sigma$]{\includegraphics[width=0.47\textwidth]{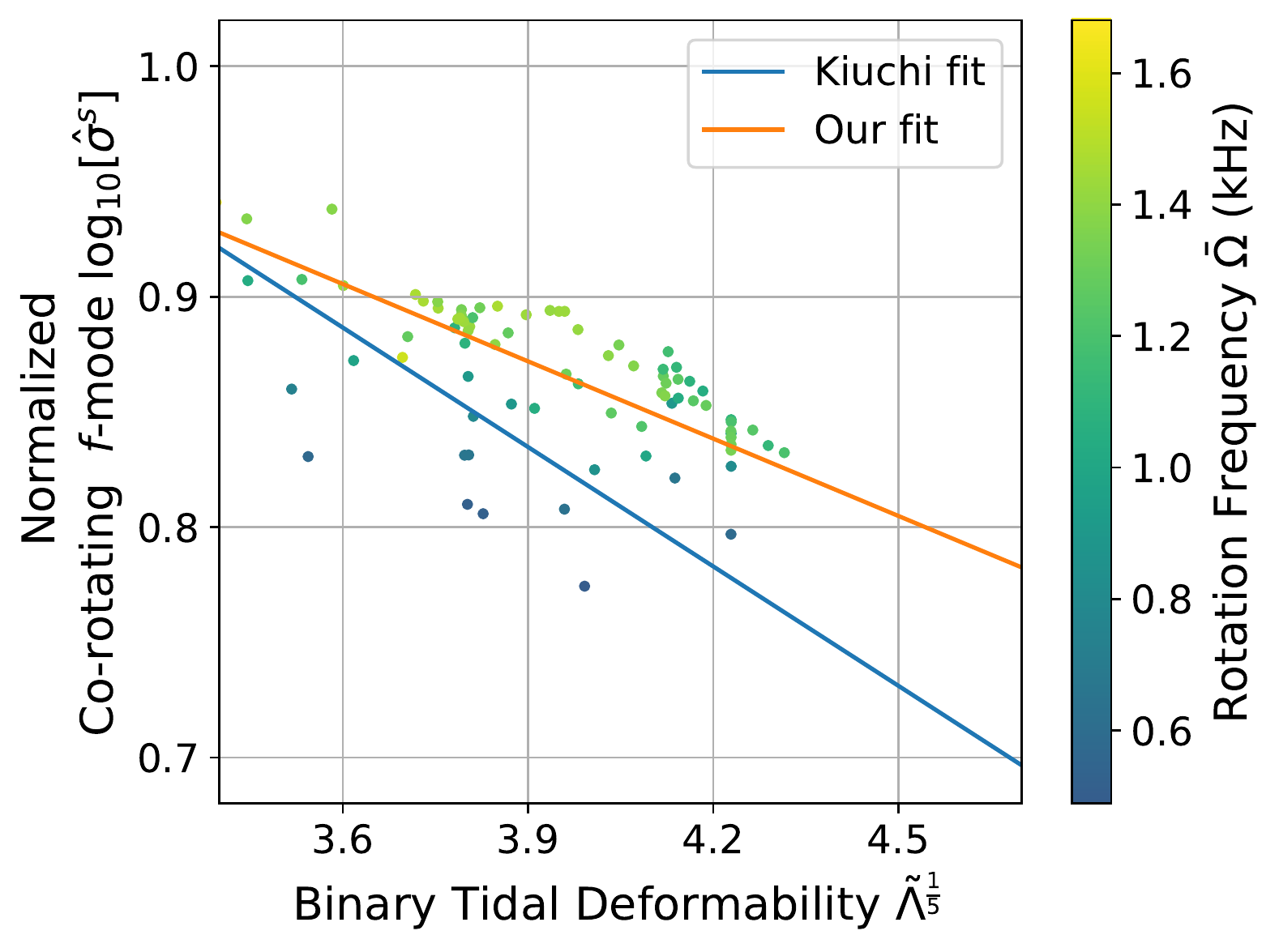}}
\subfloat[$\tilde\Lambda - \eta$]{\includegraphics[width=0.47\textwidth]{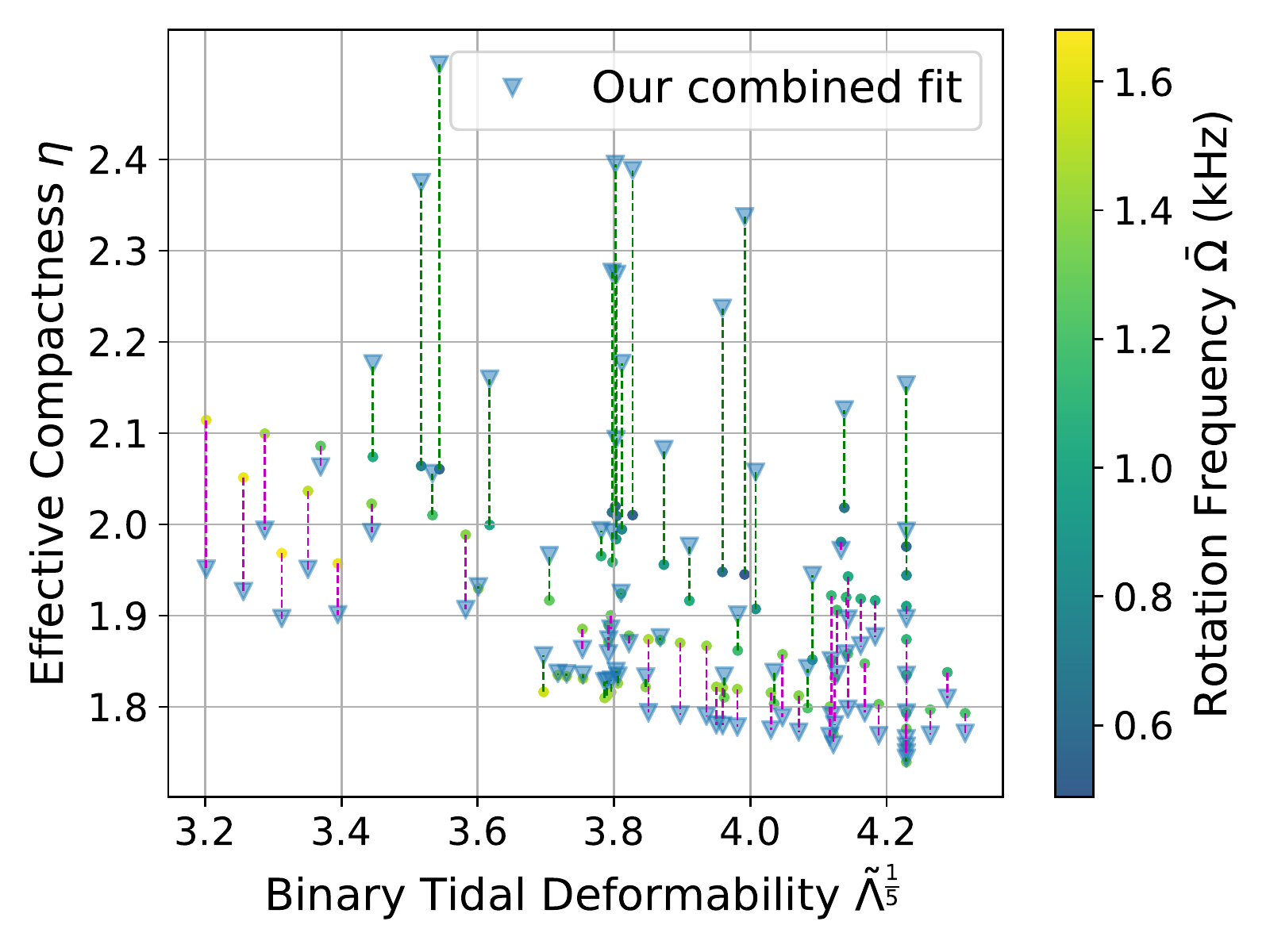}}
\caption{The $\tilde\Lambda - \hat \sigma^s$ and combined $\tilde\Lambda - \eta$ relation, using a remnant rotation threshold of $\bar \Omega = 400\si{Hz}$, for $q=1$.}
\label{fig:app-combined-400}
\end{figure*}
In general, the metric of a non-rotating neutron star is given by
\begin{equation}
ds^2 = -e^{\nu(r)} dt^2 + e^{\lambda(r)} dr^2 + r^2 d\Omega^2
\end{equation}
where the function $\lambda(r)$ fulfills the conditions
\begin{equation}
e^{\lambda(r)} = \left(1-\frac{2 m(r)}{r}\right)^{-1} \quad \text{with}\quad m(r) = \int_0^r 4\pi \bar r^2 \epsilon(\bar r) d\,\bar r 
\end{equation}
With the energy-momentum tensor of a perfect fluid
\begin{equation}
T_{\mu\nu}=(\epsilon + p)u_\mu u_\nu + p g_{\mu\nu}
\end{equation}
the metric can be determined by solving the Tolman-Oppenheimer-Volkoff (TOV) equations
\begin{equation}
\frac{d m}{d r} = 4 \pi r^2 \epsilon(r)
\end{equation}
\begin{equation}
\frac{d p}{d r} = - (\epsilon + p) \frac{m + 4 \pi r^3 p}{r(r - 2m)}
\end{equation}
\begin{equation}
\frac{d \nu}{d r} = 2 \frac{m + 4\pi r^3 p(r)}{r(r-2 m)}
\end{equation}
through integration from the center of the neutron star to its surface. We determine the surface to be at the radius $R$ at which the pressure first reaches $p(R) = 0$.

For the initial values, we set $m(0)=0$ (and thus $e^{\lambda(0)} = 1$). Given a central energy density $\epsilon(0)$, the corresponding central pressure can 
be determined by Equations~\eqref{eq:eos} and~\eqref{eq:eden_rhoc}.

Since we aim to evolve neutron stars with given gravitational masses $M_1$ and $M_2$ for a BNS of total mass $M = M_1 + M_2$, we need to specify the correct central density $\rho_c$ that results in these masses. For low to medium mass neutron stars, we use a simple bisection search to find the correct central density. Closer to the maximum mass limit, the bisection search usually fails, and we fall back to linearly searching through the full range of central densities.  
\begin{figure*}[t]
\subfloat[$\tilde\Lambda - \hat \sigma$]{\includegraphics[width=0.47\textwidth]{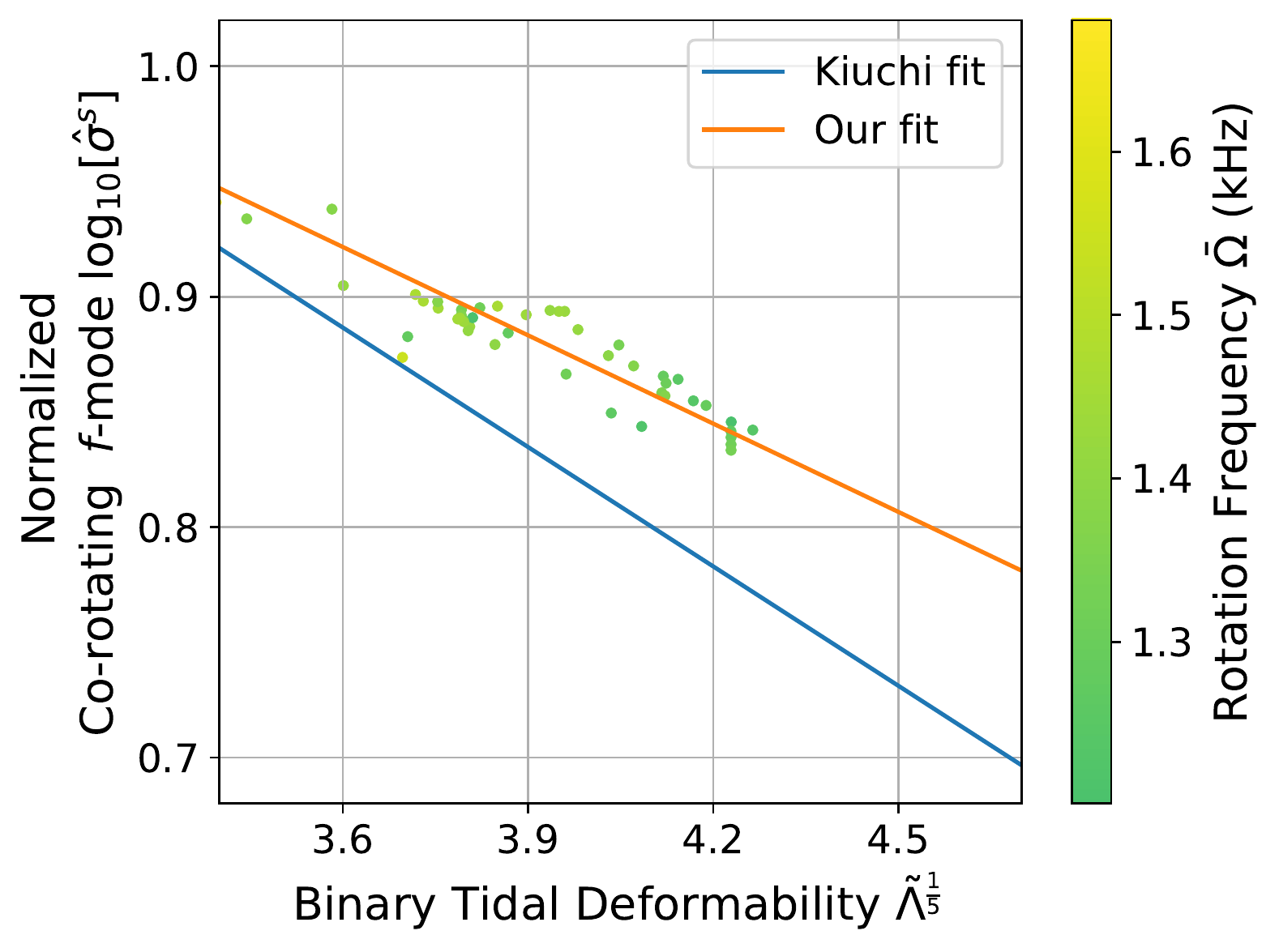}}
\subfloat[$\tilde\Lambda - \eta$]{\includegraphics[width=0.47\textwidth]{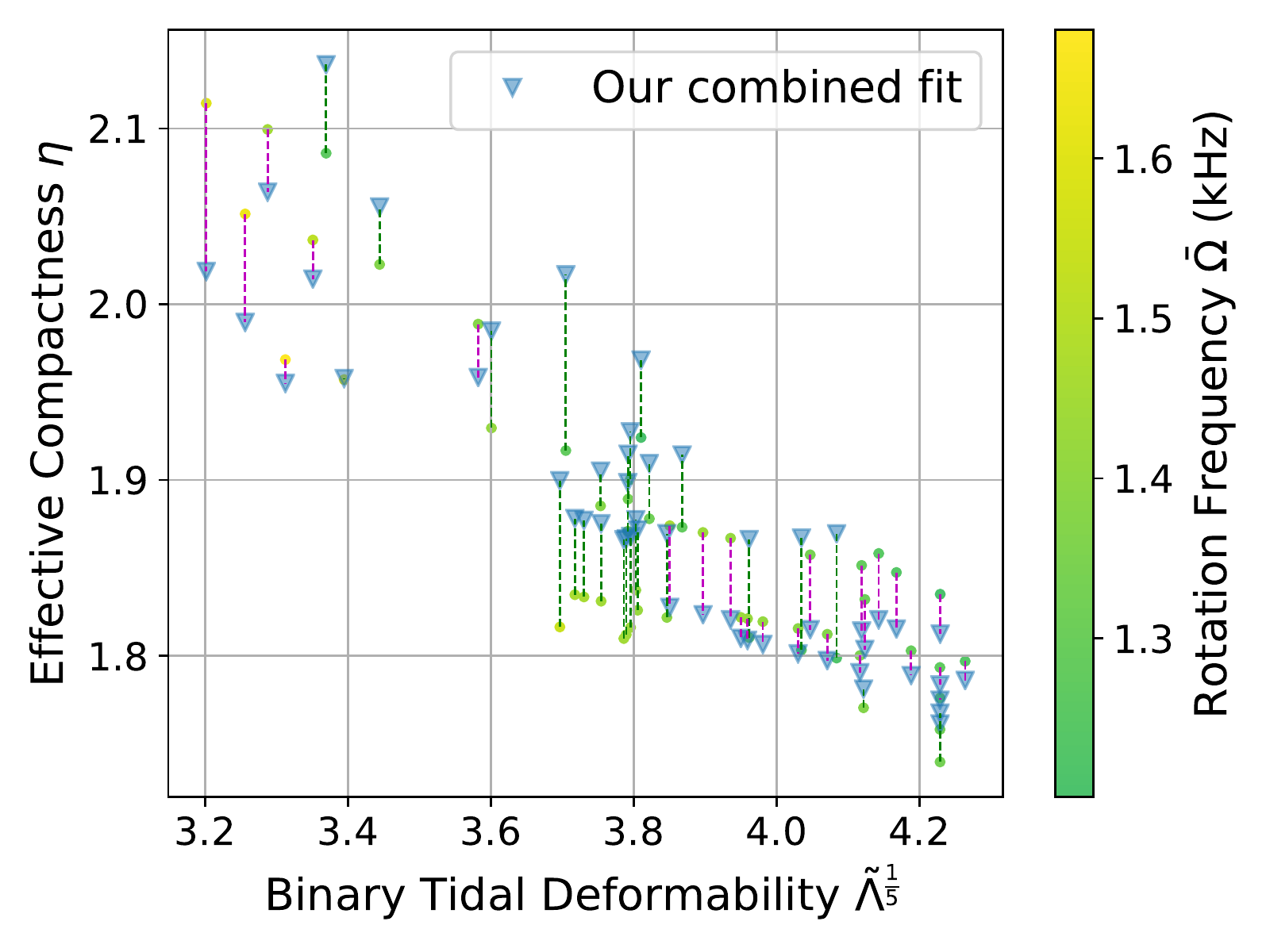}}
\caption{The $\tilde\Lambda - \hat \sigma^s$ and combined $\tilde\Lambda - \eta$ relation, using a remnant rotation threshold of $\bar \Omega = 1200\si{Hz}$, for $q=1$.}
\label{fig:app-combined-1200}
\end{figure*}

\section{Combined Relation for Different Rotation Thresholds}\label{app:combined}
Here, we present the $\tilde\Lambda - \hat \sigma^s$ relations for the rotation frequency thresholds not shown in Section~\ref{sec:combined}. The results for $\bar \Omega_{\mathrm{thr}} = 400 \si{Hz}$ are shown in Figure~\ref{fig:app-combined-400},
and the results for $\bar\Omega_{\mathrm{thr}} = 1200 \si{Hz}$ in Figure~\ref{fig:app-combined-1200}.
The coefficients for each of presented fits are listed in Table~\ref{tab:combined_res}.

\section{Direction Relations with all EoSs}\label{app:direct}
We here present the results for the direct relation considering all of the EoSs we described in Section~\ref{sec:equi}: in addition to the soft EoSs (WFF1, APR4 and SLy) we considered in the main body of the paper, we now also include the stiff EoSs (H4 and MS1). Throughout all of the presented plots, the points corresponding to the stiff EoSs will be indicated in red. 

Note that the linear, two-parameter relation with all EoSs was already discussed in Section~\ref{sec:direct_linear}. It's errors are again given in the first row of Table~\ref{tab:constraint}.
\begin{table}[t]
\sisetup{round-mode=places,round-precision=3}
\centering
\renewcommand{\arraystretch}{1.4}
\setlength\tabcolsep{1.5ex}
\begin{tabular}{c ? c | c | c}
& $\overline{E}$ &$\bar e$ & $e_{max}$ \\\specialrule{.2em}{.1em}{.1em} 
linear, two-param. & \num{0.04492761688925639} &\num{0.01675696016561328} & \num{0.07281538951794755} \\
quadratic, one-param. & \num{0.037486303958613014} & \num{0.015343700234504293} & \num{0.05944452627454945} \\ 
constr. $\Omega$ & \num{0.03491078198986463} &\num{0.0141107407142817} & \num{0.047745383705942696} 
\end{tabular}
\caption[Errors of Direct Relation between Binary Tidal Deformability and Effective Compactness]{RMSE $\bar E$, average relative error $\bar e$ and maximum relative error $e_{max}$ achieved for the relations in Equations~\eqref{eq:2-param-fit} (linear, two-param.) and \eqref{eq:quadratic-fit} (quadratic, one-param.), as well as quadratic relation with constrained rotation rates (constr. $\Omega$), for all EoSs.}
\label{tab:constraint}
\end{table}

\subsection {Quadratic, One-Parameter Relation}\label{app:direct_quad}
\begin{figure*}
    \subfloat[Effective Compactness Relation (cf. Equation~\eqref{eq:quadratic-fit})]{\includegraphics[width=0.5\textwidth]{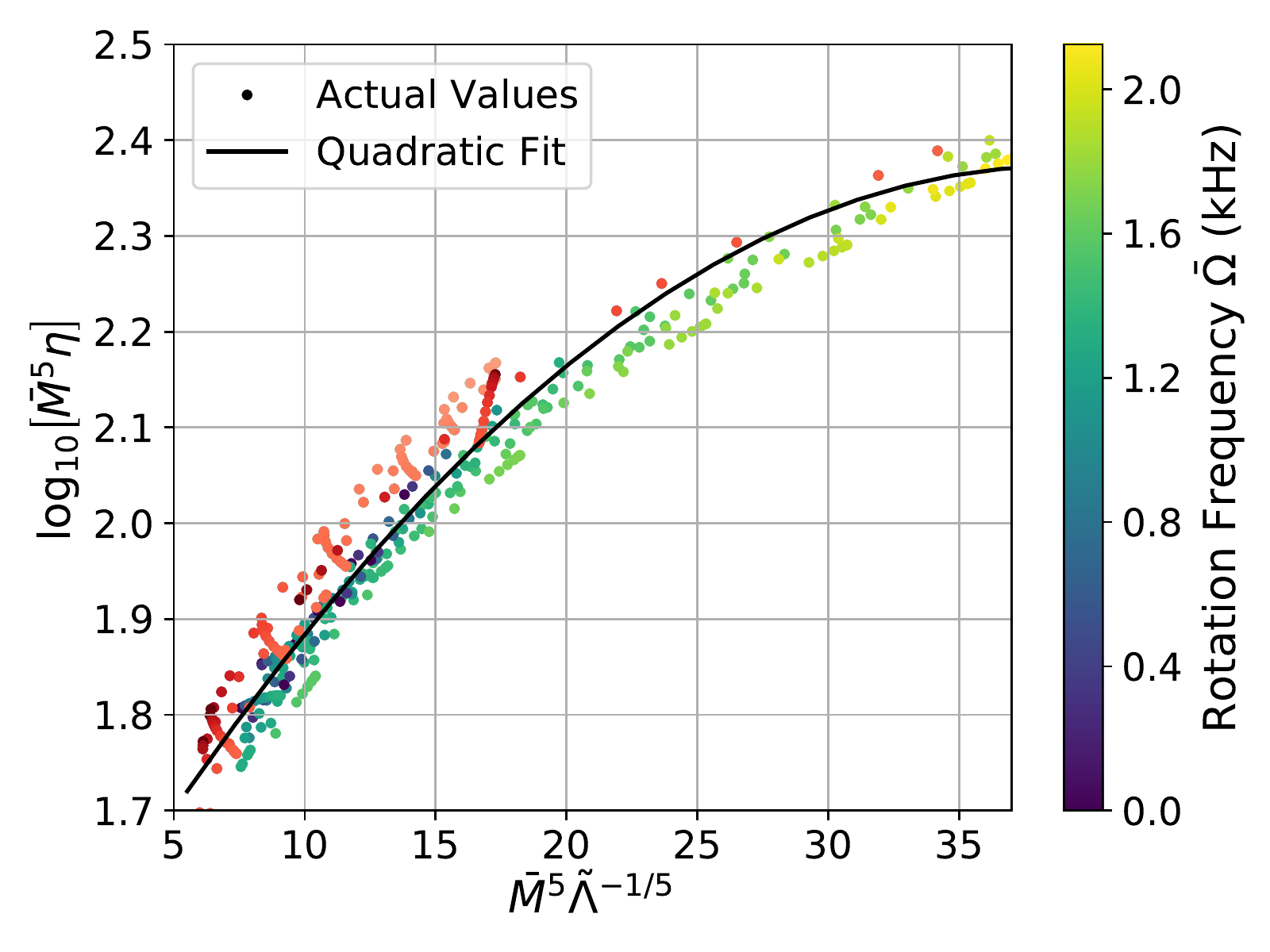}\label{fig:quadratic_lambda_eta_fit_all}}
    \subfloat[Compactness Relation (cf. Equation~\eqref{eq:compactness-fit})]{\includegraphics[width=0.5\textwidth]{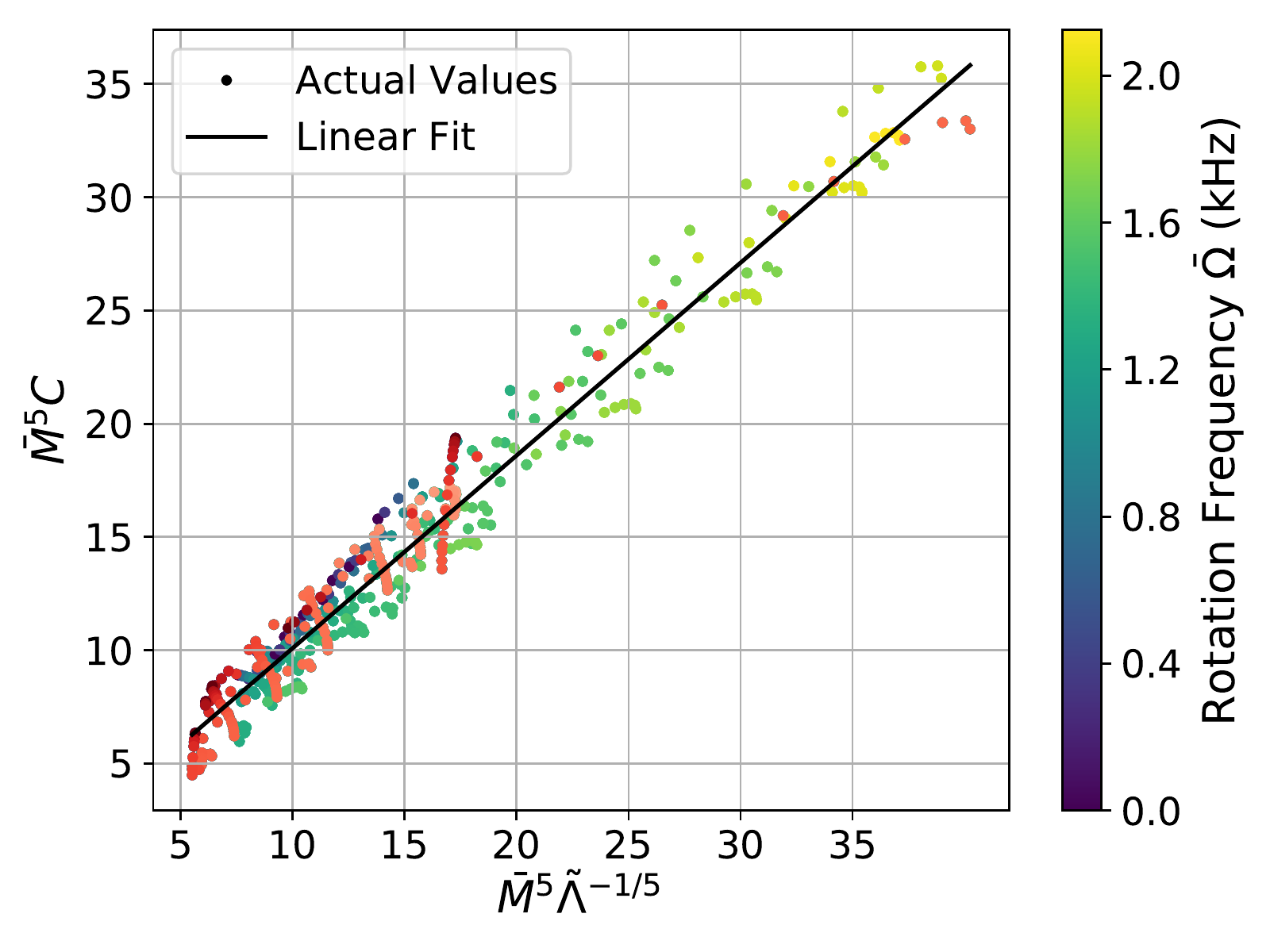}\label{fig:compactness_lambda_fit}}
    \caption{The best fits for the one-parameter effective compactness and compactness relations, for $q=1$, considering soft and stiff EoSs. Stiff EoSs are indicated in red}
\end{figure*}

We first revisit the quadratic, one-parameter relation discussed in Section~\ref{sec:direct_quadratic}, now with all EoSs. The relation is illustrated in Figure~\ref{fig:quadratic_lambda_eta_fit_all}. The coefficients of the best fit are given in Table~\ref{tab:coeff-quadratic} and the resulting errors are given in the second row of Table~\ref{tab:constraint}. We go from an average relative error of $0.9\%$ for the case with only soft EoSs, to now $1.5\%$. 

\begin{table}[t!]
\sisetup{round-mode=places,round-precision=3,scientific-notation=true}
\centering
\setlength\tabcolsep{1  ex}
\renewcommand{\arraystretch}{1.4}
\begin{tabular}{c ? c | c | c }
 $i$ & 2 & 1 & 0 \\\specialrule{.2em}{.1em}{.1em}
 $a_i$ & \num{-0.00014865} & \num{0.00029043} & \num{-0.00072965} \\
$b_i$ & \num{0.00603859} & \num{-0.01180149} & \num{0.05142265} \\
$c_i$ & \num{-0.00783046} & \num{0.01520246} & \num{1.47895638}
\end{tabular}
\caption{Coefficients of the quadratic, one-parameter relation in Equation~\eqref{eq:quadratic-fit}, considering soft and stiff EoSs. The fit is illustrated in Figure~\ref{fig:quadratic_lambda_eta_fit_all}.}
\label{tab:coeff-quadratic}
\end{table}
\begin{table}[b]
\sisetup{round-mode=places,round-precision=3}
\centering
\setlength\tabcolsep{1ex}
\renewcommand{\arraystretch}{1.4}
\begin{tabular}{c ? c | c | c }
$i$ & 2 & 1 & 0 \\\specialrule{.2em}{.1em}{.1em}
 $a_i$ & \num{0.12272452} & \num{-0.24011706} & \num{0.96812734} \\
 $b_i$ & \num{-0.29213048} & \num{0.57180147} & \num{1.30624574} \\
\end{tabular}{}
\caption{Coefficients of the linear, one-parameter relation in Equation~\eqref{eq:compactness-fit}, considering soft and stiff EoSs. This fit is illustrated in Figure~\ref{fig:compactness_lambda_fit}.}
\label{tab:coeff-compactness}
\end{table}

\subsection{Compactness Relation for all EoSs}\label{app:compactness}
We here revisit the compactness relation discussed in Section~\ref{sec:compactness}, now for all EoSs. The relation is illustrated in Figure~\ref{fig:compactness_lambda_fit}, with the corresponding coefficients given in Table~\ref{tab:coeff-compactness}. 
The relation achieves an average relative error of $8.8\%$ once we introduce the stiff EoSs, which again is slightly worse than for the case with only soft EoSs, where we achieved $7.6\%$. 

\subsection{Relations with Mass Loss for all EoSs}\label{app:mass_loss_all}
\begin{figure*}
\subfloat[$\Delta M = 0.1 M_\odot$]{\includegraphics[width=0.5\textwidth]{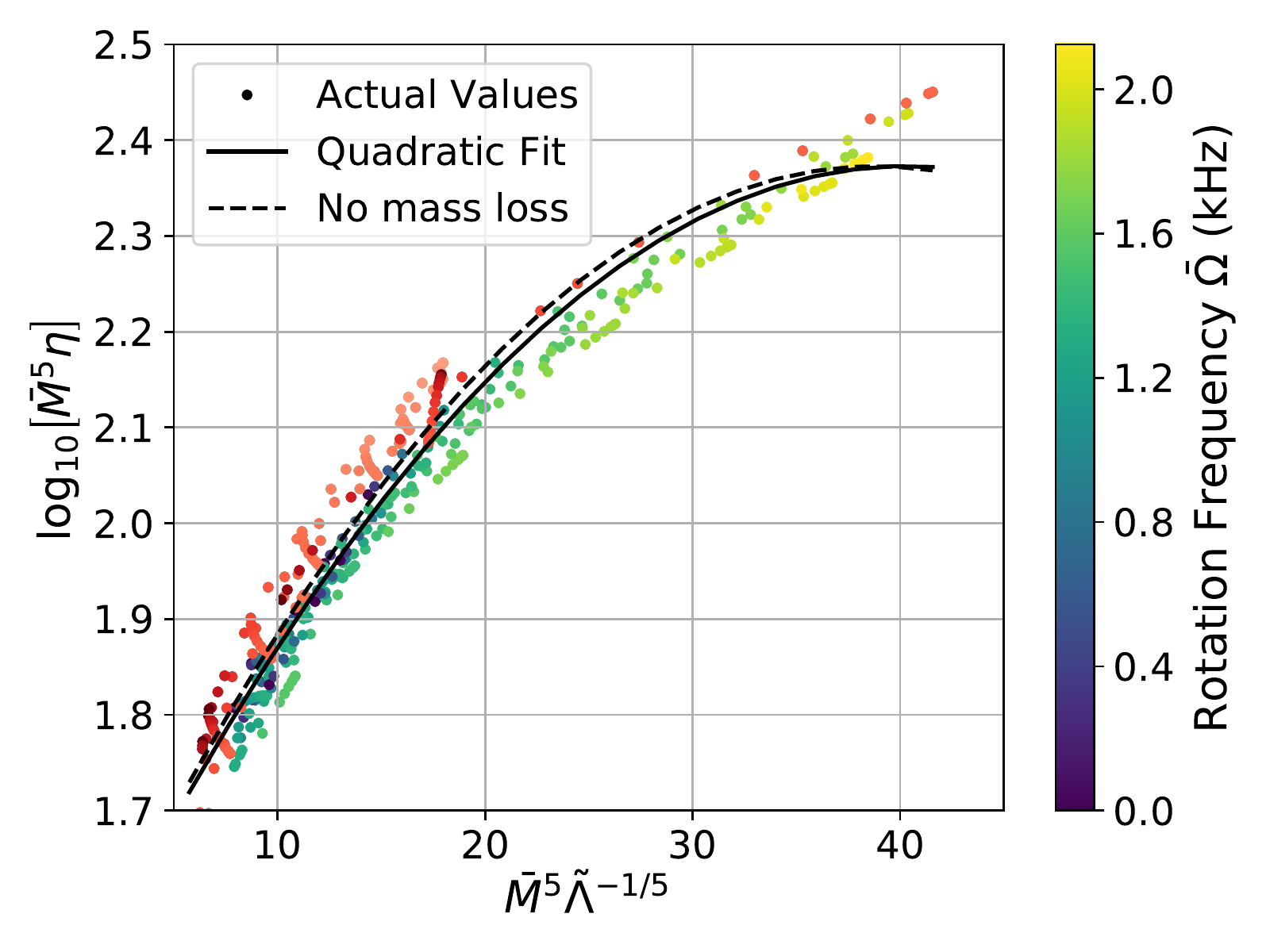}}
\subfloat[$\Delta M = 0.2 M_\odot$]{\includegraphics[width=0.5\textwidth]{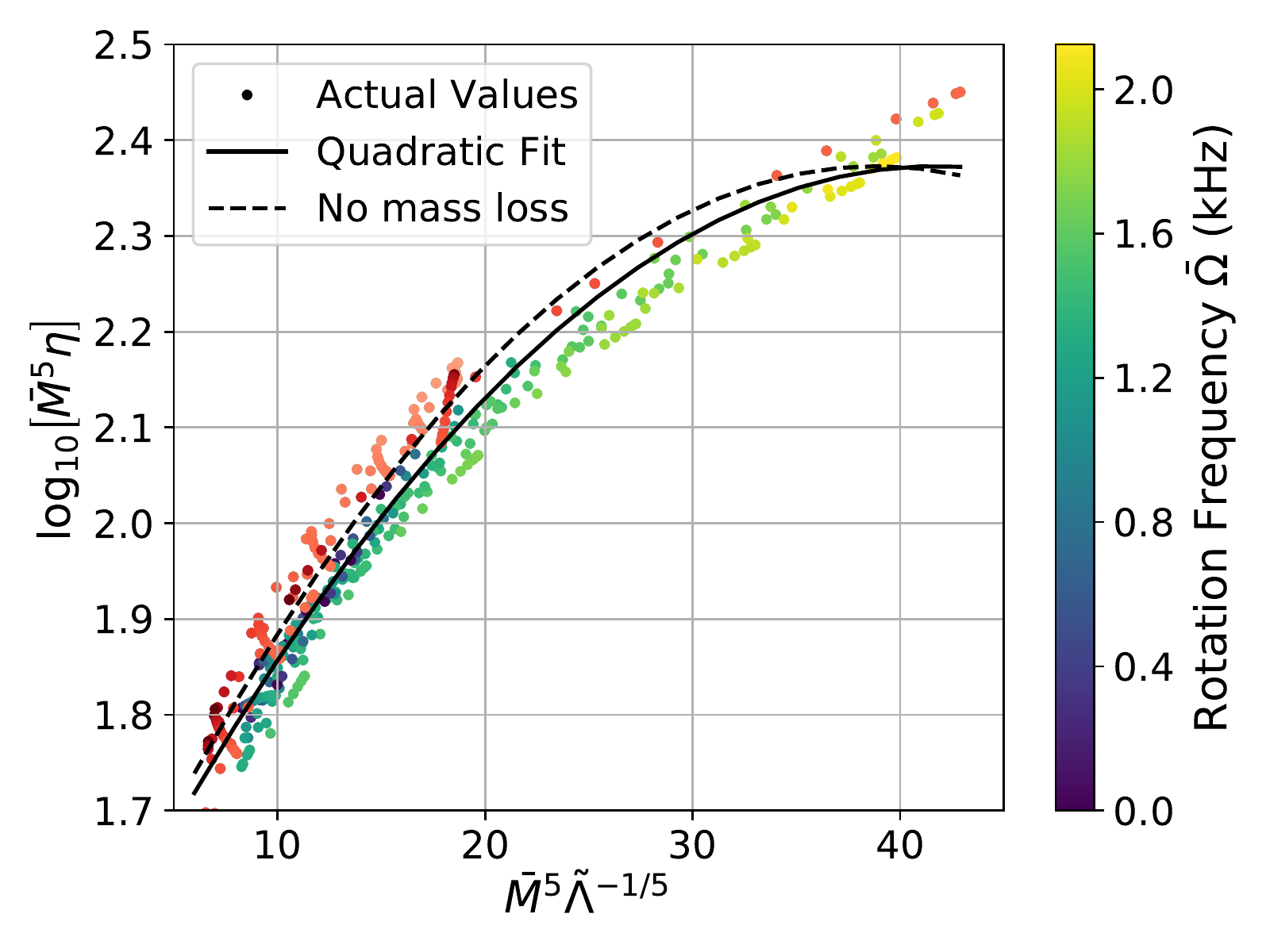}}
\caption{The quadratic fit (cf. Equation~\eqref{eq:quadratic-fit}) between $\bar M^{5} \tilde \Lambda^{-\frac{1}{5}}$ and $\log_{10}\left[\bar M^5 \eta\right]$ for $q=1$, with a $\Delta M$ baryon mass loss, considering all EoSs. The dashed line represents the original fit.}
\label{fig:app-01_mass_loss-quad_all_EoS}
\end{figure*}

We finally also investigate the impact of a non-negligible mass going from the pre-merger stars to the long-lived remnant while considering soft and stiff EoSs. In Figure~\ref{fig:app-01_mass_loss-quad_all_EoS}, we show the data points and new best fits for the quadratic, one-parameter relation, together with the original fit (without mass loss) represented as a dashed line. 

In Table~\ref{tab:errors_mass_loss_comparison_quad_all} we list the errors of the two new fits, and compare them to the old fit without mass loss. Clearly, the new fits are exactly as accurate as the old relations, similar to what we observed in the main part of the paper for the case of only soft EoSs.

In Table~\ref{tab:errors_02_mass_loss_quadratic_all}, we show how accurate our original relation (without mass loss) is when applied to mergers with mass loss. As with the case with only soft EoSs that we considered in the main body of the paper, the case with all EoSs also incurs a small decrease in the accuracy of our original relation with increasing mass loss, in the range of $0.5 \%$ for the average relative error.

Consequently, the behavior of our relations when considering a non-negligible mass loss is identical, independent of whether we consider only soft or all EoSs.

\begin{table}
\sisetup{round-mode=places,round-precision=5}
\centering
\renewcommand{\arraystretch}{1.4}
\setlength\tabcolsep{1.5ex}
\begin{tabular}{c ? c | c | c}
$\Delta M$ & $\overline{E}$ &$\bar e$ & $e_{max}$ \\\specialrule{.2em}{.1em}{.1em} 
0 & \num{0.037486303958613014} & \num{0.015343700234504293} & \num{0.05944452627454945} \\ 
$0.1 M_\odot$ & \num{0.03750518616593095} & \num{0.015388815891021326} & \num{0.058853315023818575} \\
$0.2 M_\odot$ & \num{0.03751823286002482} & \num{0.015431586578407984} & \num{0.05824617927765413} \\ 
\end{tabular}
\caption[Errors of Direct Relation between Binary Tidal Deformability and Effective Compactness]{RMSE $\bar E$, average relative error $\bar e$ and maximum relative error $e_{max}$ achieved for the quadratic, one-parameter relation (cf. Equation~\eqref{eq:quadratic-fit}), with and without mass loss. Considering all EoSs.}
\label{tab:errors_mass_loss_comparison_quad_all}
\end{table}

\begin{table}
\sisetup{round-mode=places,round-precision=3}
\centering
\renewcommand{\arraystretch}{1.4}
\setlength\tabcolsep{1.5ex}
\begin{tabular}{ c ? c | c | c}
 Mass Loss & $\overline{E}$ &$\bar e$ & $e_{max}$ \\\specialrule{.2em}{.1em}{.1em} 
0 & \num{0.037486303958613014} &\num{0.015343700234504293} & \num{0.05944452627454945} \\
$0.1 M_\odot$ & \num{0.04010544747707366} & \num{0.017073769592507265} & \num{0.06513330386174915} \\ 
$0.2 M_\odot$& \num{0.04694121479507198} &\num{0.020137508455363433} & \num{0.07081128310597225} 
\end{tabular}
\caption{RMSE $\bar E$, average relative error $\bar e$ and maximum relative error $e_{max}$ achieved by our original quadratic, one-parameter fit (cf. Equations~\eqref{eq:quadratic-fit}), for mergers with $0$, $0.1 M_\odot$ and $0.2 M_\odot$ baryon mass loss. Considering all EoSs}
\label{tab:errors_02_mass_loss_quadratic_all}
\end{table}

\onecolumngrid

\end{document}